\documentclass{article}

\usepackage{latexsym}

\usepackage{amssymb}
\def\prfe{\hspace*{\fill} $\Box$

\smallskip \noindent}

\newcommand{\fzero}{f^{\mathrm{in}}}

\newtheorem{Theorem}{Theorem}
\newtheorem{Proposition}{Proposition}
\newtheorem{Lemma}{Lemma}                                                          
\newtheorem{Definition}{Definition} 
\newtheorem{Corollary}{Corollary}

\DeclareFontFamily{OT1}{rsfs}{}                                                 
\DeclareFontShape{OT1}{rsfs}{m}{n}{ <-7> rsfs5 <7-10> rsfs7                     
<10->rsfs10}{}                                                                  
\DeclareMathAlphabet{\mycal}{OT1}{rsfs}{m}{n}

\renewcommand{\theequation}{\thesection.\arabic{equation}}

\title{Global Small Solutions of the Vlasov-Maxwell System in the Absence of Incoming Radiation}
\author{Simone Calogero\\[0.5cm]
Max Planck Institut f\"ur Gravitationphysik\\
Albert Einstein Institut Am M\"uhlenberg 1\\
14476 Golm bei Potsdam, Germany\\
E-mail: simcal1@aei-potsdam.mpg.de}
\date { }
                                                                              
\begin{document} 
\maketitle

\begin{abstract}
We consider a modified version of the Vlasov-Maxwell system in which the usual Maxwell fields are replaced by their retarded parts. We
show that solutions of this modified system exist globally for a small initial density of particles and that they describe a system without 
incoming radiation. 

\end{abstract}

\section{Introduction and main result}\label{introduction}
\setcounter{equation}{0}
The relativistic Vlasov-Maxwell system (RVM) models the dynamics of a plasma consisting of a large number of
charged particles under the assumption that the particles interact only by the electrodynamic forces that the fields generate collectively. In
particular, collisions between particles and external forces are assumed to be negligible.

Examples of physical systems which are thought to be well-modelled by RVM are the solar wind and the ionosphere.

Given the huge number of particles which form the plasma it should be hopeless to attempt to describe the state of the 
plasma by looking at the position and the velocity of each individual particle. Therefore a statistical description of the matter is 
needed. In the framework of kinetic theory the microscopic state of the plasma is described by specifying a distribution function in the phase space for 
each species of particle.  
Let us assume for simplicity that the plasma consists of a single species of particle with unit mass and charge and set also the speed of
light equal to one (i.e. $c=1$). We denote by $f(t,x,p)$ 
the probability density to find a particle at time $t$ at position $x$ with momentum $p$, where 
$(t,x,p)\in\mathbb{R}_{t}\times\mathbb{R}^{3}_{x}\times\mathbb{R}^{3}_{p}$.
Clearly, $f\geq 0$. The charge 
density and the current density of the plasma are given respectively by
\begin{equation}\label{sources}
\rho(t,x)=\int_{\mathbb{R}^{3}} dp\,f(t,x,p),\quad
j(t,x)=\int_{\mathbb{R}^{3}} dp\,\widehat{p}\,f(t,x,p),
\end{equation}
where we denoted by $\widehat{p}$ the relativistic velocity of a particle with momentum $p$, that is
\begin{equation}\label{velocity}
\widehat{p}=\frac{p}{\sqrt{1+|p|^{2}}}.
\end{equation}

The electromagnetic field $(E,B)$ generated by the plasma solves the Maxwell equations
\begin{equation}
\left\{\begin{array}{ll}\partial_{t} E=\partial_{x}\wedge B- 4\pi j,& \quad \partial_{x}\cdot E=4\pi\rho,\\
\partial_{t} B=-\partial_{x}\wedge E,& \quad\partial_{x}\cdot B=0.\end{array}\right.
\end{equation}
The system is closed by requiring that $f$ be a solution of the Vlasov continuity equation
 \begin{equation}\label{vlasov}
\partial_{t}f+\widehat{p}\cdot\partial_{x}f+(E+\widehat{p}\wedge B)
\cdot\partial_{p}f=0.
\end{equation}

The RVM system consists of the set of equations (\ref{sources})--(\ref{vlasov}). A short survey on the initial value problem for this system will be given at the end of this
introduction.
For later convenience we recall here the definition of the total energy of a solution of RVM, which is  
$\mathcal{E}_{\mathrm{tot}}(t)=\mathcal{E}_{\mathrm{kin}}(t)+\mathcal{E}_{\mathrm{field}}(t)$,
where $\mathcal{E}_{\mathrm{kin}}(t)$ is the kinetic energy of the particles,
\begin{displaymath}
\mathcal{E}_{\mathrm{kin}}(t)=\int dx\int dp\,\sqrt{1+|p|^{2}}f(t,x,p) 
\end{displaymath}
and $\mathcal{E}_{\mathrm{field}}(t)$ is the field energy,
\begin{displaymath}
\mathcal{E}_{\mathrm{field}}(t)=\frac{1}{2}\int dx\, (|E(t,x)|^{2}+|B(t,x)|^{2}).
\end{displaymath}
(In the previous definitions it is understood that the integrals are 
extended over $\mathbb{R}^{3}$). For smooth solutions of RVM the total energy is finite and conserved for all times provided it is finite
at the time $t=0$ (cf. \cite{GS1}).

In this paper we are interested in those solutions of RVM which are characterized by the property of being isolated from 
\textit{incoming radiation}. Let us first discuss these solutions heuristically and then we will give their precise definition. 

The radiation is defined as the part of the electromagnetic field which carries energy to  
\textit{null infinity}, that is to that part of the infinity of the Minkowski space which is reached along the null and 
asymptotically null curves. The null infinity is distinguished in \textit{future} null infinity, which is reached in the limit
$t\to +\infty$, $|x|\to +\infty$, at constant \textit{retarded} time $u=t-|x|$, and \textit{past} null infinity, which
is reached in the limit $t\to -\infty$, $|x|\to +\infty$, now at constant \textit{advanced} time,
$v=t+|x|$. Correspondingly one defines \textit{outgoing} and \textit{incoming} radiation to be the part of the electromagnetic field
which propagates energy to future and past null infinity respectively. 

Since RVM is symmetric with respect to the
transformation $t\to -t$ (time reflection\footnote{Namely, if $(f(t,x,p),E(t,x),B(t,x))$ is a solution, then $(f(-t,x,-p), E(-t,x),
-B(-t,x))$ gives a new solution of RVM.}), this system will contain in general outgoing as well as incoming radiation.      
In order to give a precise definition of solutions of RVM which do not contain incoming radiation, let us consider the energy 
$\mathcal{E}_{\mathrm{in}}(v_{1},v_{2})$ carried by the field
to past null infinity in the interval $[v_{1},v_{2}]$ of the advanced time. This quantity can be formally calculated by the limit
\[
\mathcal{E}_{\mathrm{in}}(v_{1},v_{2})=
-\lim_{r\to +\infty}\int_{v_{1}}^{v_{2}}dv\int_{|x|=r}dx\,(E\wedge B)\cdot \omega_{\mid t=v-r},
\]
where $\omega=x/|x|$ and $E\wedge B$ is the Poynting vector (the minus sign comes from the convention to consider positive the flux of energy flowing in
onto the system). We will say that a solution of RVM is isolated from incoming radiation if $\mathcal{E}_{\mathrm{in}}(v_{1},v_{2})=0$, for all
$v_{1}, v_{2}\in\mathbb{R}$.

In this paper we are mainly concerned with the question whether the solutions of RVM isolated from incoming radiation are represented by 
the retarded solution of the equations.
For this purpose we restrict ourselves to consider the system 
\begin{equation}\label{vlasov2}
\partial_{t}f+\widehat{p}\cdot\partial_{x}f+(E_{\mathrm{ret}}+\widehat{p}\wedge B_{\mathrm{ret}})
\cdot\partial_{p}f=0,
\end{equation}
\begin{equation}\label{eletfield}
E_{\mathrm{ret}}(t,x)=-\int\frac{dy}{|x-y|}(\partial_{x}\rho+\partial_{t}j)(t-|x-y|,y),
\end{equation}
\begin{equation}\label{magfield}
B_{\mathrm{ret}}(t,x)=\int\frac{dy}{|x-y|}(\partial_{x}\wedge j)(t-|x-y|,y),
\end{equation}
where $\rho$ and $j$ are defined by (\ref{sources}). 
We will refer to the system (\ref{vlasov2})--(\ref{magfield}) as the \textit{retarded}
relativistic Vlasov-Maxwell system, or RVM$_{\mathrm{ret}}$ for short.

Let us briefly comment in which sense the solutions of RVM$_{\mathrm{ret}}$ have to be considered as solutions of RVM. Assume first that 
$f_{\mathrm{ret}}$ is a $C^{1}$ global solution of RVM$_{\mathrm{ret}}$ and that also $(E_{\mathrm{ret}},B_{\mathrm{ret}})$ is $C^{1}$. 
By means of (\ref{vlasov2}), 
$\rho$ and $j$ satisfy the continuity equation, $\partial_{t}\rho+\partial_{x}\cdot j=0$, and therefore the retarded field \textit{is}
a solution of the Maxwell equations. Thus, $(f_{\mathrm{ret}}, E_{\mathrm{ret}}, B_{\mathrm{ret}})$ is a solution of RVM. The same is true if 
$f_{\mathrm{ret}}$ is a
\textit{semiglobal} solution of RVM$_{\mathrm{ret}}$, i.e. a solution defined for $t\in (-\infty,T]$, where $T\in\mathbb{R}$. 
However it is clear that there is no meaningful notion of local solutions of RVM$_{\mathrm{ret}}$. For the retarded
field at a point $(t,x)$ is obtained by integration over the \textit{whole} past light cone with vertex in $(t,x)$ (no initial data
for the field are imposed) and so the field at time $t$ is determined if and only if a solution has been constructed in the
interval $(-\infty,t]$. 

We can now state the main result of this paper. This is a global existence and uniqueness theorem for small data of solutions of 
RVM$_{\mathrm{ret}}$ which we also show to be isolated from incoming radiation in the sense specified above. 
\begin{Theorem}\label{main}
Let $\fzero(x,p)\geq 0$ be given in $C_{0}^{2}(\mathbb{R}^{3}_{x}\times\mathbb{R}^{3}_{p})$ and $R>0$ such that 
$\fzero(x,p)=0$ for $|x|^{2}+|p|^{2}\geq R^{2}$. 
Define 
\[
\Delta=\sum_{|\mu|=0}^{2}\|\nabla^{\mu}\fzero\|_{\infty},
\] 
where $\mu\in\mathbb{N}^{6}$ is a multi-index.
Then there exists a constant $\varepsilon >0$  depending only on $R$ such that for $\Delta\leq\varepsilon$,
RVM$_{\mathrm{ret}}$ has a unique $C^{1}$ global solution $f_{\mathrm{ret}}$ satisfying $f_{\mathrm{ret}}(0,x,p)=\fzero(x,p)$.
Moreover $(E_{\mathrm{ret}},B_{\mathrm{ret}})\in C^{1}(\mathbb{R}_{t}\times\mathbb{R}^{3}_{x})$ and there exists a positive constant $C=C(R)$ such
that the field satisfies the following estimates for all $(t,x)\in\mathbb{R}_{t}\times\mathbb{R}^{3}_{x}$: 
\begin{eqnarray}
|E_{\mathrm{ret}}(t,x)|+|B_{\mathrm{ret}}(t,x)| &\le& C\Delta (1+|t|+|x|)^{-1}(1+|t-|x||)^{-1},\label{estimate1}\\
|DE_{\mathrm{ret}}(t,x)|+|DB_{\mathrm{ret}}(t,x)| &\le& C\Delta (1+|t|+|x|)^{-1}(1+|t-|x||)^{-7/4},\label{estimate2}
\end{eqnarray}
where $D$ denotes any first order derivative.
Moreover $(f_{\mathrm{ret}},E_{\mathrm{ret}},B_{\mathrm{ret}})$ is the unique solution of RVM which satisfies (\ref{estimate1}), (\ref{estimate2}) and 
$f_{\mathrm{ret}}(0,x,p)=\fzero(x,p)$. 
\end{Theorem}

The uniqueness assertion of theorem \ref{main} will be made more precise in proposition \ref{unique} below. 
The fact that the solution of theorem 1 is isolated from incoming radiation is a consequence of the estimate (\ref{estimate1}).

This paper is organized as follows. In section \ref{preliminary} we recall a few facts on RVM which will be needed in the sequel. 
In section \ref{estimates} we prove the
main estimates on the retarded field and the uniqueness part of theorem \ref{main}. The existence part is proved in section \ref{existence}.
An appendix is devoted to the proof of two technical lemmas.
 
To conclude this introduction we mention some important results on the initial value problem for RVM. 
Existence of a unique solution for a short time has been proved in \cite{W}. 
A unique global solution is shown to exist in \cite{GS2} under the \textit{a priori}  assumption that there exists a function $\beta\in
C^{0}(\mathbb{R})$  
such that $\mathcal{P}(t)\leq \beta(t),\, \forall t\in\mathbb{R}$,
where $\mathcal{P}(t)$ denotes the maximum momentum of the particles up to the time $t$, i.e.:
\[
\mathcal{P}(t)=\sup_{0\leq s\leq t}\{|p|:f(s,x,p)\neq 0,\textnormal{ for some }x\in\mathbb{R}^{3}\}.
\]
A different proof of this result based on the Fourier transform was given recently in \cite{KS}.

The result in \cite{GS2} was applied to prove global existence and uniqueness under different smallness assumptions on the initial
data (cf. \cite{GS4,GSh1,R}) and for arbitrarily large data in two space dimensions (i.e. $x\in\mathbb{R}^{2}$)
in \cite{GSh2}. Existence, but not uniqueness, of global weak solutions is proved in \cite{DipLio}.

The non-relativistic limit of RVM is the Vlasov-Poisson system (VP). For a single species of particles with unit positive charge and 
mass the VP system is given by
\[
\partial_{t}f+v\cdot\partial_{x}f+\partial_{x}U\cdot\partial_{v}f=0,
\]
\[
\Delta U=4\pi\rho,
\]
where $U$ is the electrostatic potential, $v$ the classical velocity of the particles, $f=f(t,x,v)$ and $\rho=\int dv\, f$.
The initial value problem for VP has been proved to be correctly set for general initial data in \cite{LP,P} (cf. also \cite{R2,Sch}) 
and the convergence of solutions of RVM to solutions of VP, when the speed of light tends to infinity, has been established rigorously 
in \cite{Sh}.
The \textit{a priori} estimates proved in \cite{IR} show that the solutions of VP do not contain radiation. In order to measure an energy lost 
to infinity for VP (in a non-relativistic sense, i.e.
at \textit{spacelike} infinity) an extra dipole term has to be added into the equations (cf. \cite{KR}).

\section{Preliminary results}\label{preliminary}
\setcounter{equation}{0}
In this section we recall some well-known results on RVM which will be used later on. We start by fixing a bit of notation.
The symbol $T$ will denote the \textit{free transport} operator, that is
\[
T=\partial_{t}+\widehat{p}\cdot\partial_{x}.
\]
We denote by $C$ a generic constant which may change from line to line but which depends only on $R$.
If a constant depends on $R$ and on other parameters, it will be denoted by $C_{*}$. The partial derivative with respect to $x_{i}$
$(i=1,2,3)$ will be denoted by $\partial_{x_{i}}$, while any derivative of order $k$  with respect to $t$ and/or $x$ will be denoted by 
$D^{k}$ (namely, $Dg=\partial_{t}g$ or $\partial_{x_{i}}g$,
$D^{2}g=\partial^{a}_{t}\partial_{x_{i}}^{b}\partial_{x_{j}}^{c}g$, $a+b+c=2$ and so on, with the convention $D^{0}g=g$).
The $L^{\infty}$ norm of a function $g(x_{1},...,x_{n})$ with respect to the variables $x_{k+1},...,x_{n}$ will be denoted by 
$\|g(x_{1},...,x_{k})\|_{\infty}$. The $L^p$ norm is denoted by $\|\,\cdot\,\|_{L^p}$. The notation $\|\,\cdot\,\|_w$ is used for the norm
defined in section \ref{existence} below (cf. (3.4)). We also set $F=(E,B)$.

The Vlasov equation can be reduced to a system of ordinary differential equations by using the method of characteristics. 
Consider the following ``initial'' value problem for the function $(X,P):\mathbb{R}_{s}\to \mathbb{R}^{6}$:
\begin{eqnarray}
&&\frac{dX}{ds}=\widehat{P},\label{charac1}\\
&&\frac{dP}{ds}=E(s,X)+\widehat{P}\wedge B(s,X),\label{charac2}\\
&&(X(t),P(t))=(x,p).
\end{eqnarray}
Let $(X(s,t,x,p),P(s,t,x,p))$ denote the solution of the previous problem (sometimes it will be denoted by $(X(s),P(s))$ for short). 
Then the solution of the Vlasov equation is given by
\begin{equation}\label{solvlasov}
f(t,x,p)=\fzero(X(0,t,x,p),P(0,t,x,p)).
\end{equation}
By (\ref{solvlasov}), $\textnormal{supp}\, [f(t)]=\{(x,p):f(t,x,p)\neq 0\}\subseteq\Xi(t)$ where
\[
\Xi(t)=\{ (x,p)\in \mathbb{R}^{3}_{x}\times\mathbb{R}^{3}_{p} \textrm{ s.t. } |X(0,t,x,p)|^{2}+|P(0,t,x,p)|^{2}\leq R^{2} \}.
\] 
Moreover, since the characteristics flow preserves the Lebesgue measure, the $L^{p}$ norm in phase space of the particle density 
is conserved:
\begin{equation}
\|f(t)\|_{L^{p}}=\|\fzero\|_{L^{p}}, \quad\forall\, 1\leq p\leq\infty,\, t\in\mathbb{R}.
\end{equation}

We also recall the following
\begin{Definition} A solution $(f,F)$ of RVM is said to satisfy the ``Free Streaming Condition'' (FSC) 
with respect to the constant $\eta>0$ if there exists $\alpha>\frac{1}{2}$ such that 
\begin{eqnarray*}
|F(t,x)|&\leq& \eta (1+|t|+|x|)^{-\alpha} (1+R+|t|-|x|)^{-\alpha},\\
|\partial_{x}F(t,x)|&\leq& \eta (1+|t|+|x|)^{-\alpha} (1+R+|t|-|x|)^{-\alpha-1},
\end{eqnarray*} 
for $t\in\mathbb{R}$ and $|x|\leq R+|t|$.
\end{Definition}

The following lemma contains some estimates on the characteristics which are due to FSC. 
\begin{Lemma}\label{estimateschar}
There exists a constant $\eta_{0}>0$ such that if $(f,F)$ is a $C^{1}$ solution of RVM which satisfies FSC 
with respect to $\eta\leq\eta_{0}$, then for all $(x,p)\in\Xi(t)$ and $t\in\mathbb{R}$:
\begin{equation}\label{estimP}
\mathcal{P}(t)\leq 2R,
\end{equation}
\begin{equation}\label{estimchar1}
|\partial_{x}(X,P)(0,t,x,p)|\leq C,
\end{equation}
\begin{equation}\label{estimchar2}
|\partial_{p}(X,P)(0,t,x,p)|\leq C(1+|t|).
\end{equation}
Moreover for all $(x,p_{i})\in \Xi(t)$ (i=1,2) and $t\in\mathbb{R}$:
\begin{equation}\label{estimsupp}
|X(0,t,x,p_{1})-X(0,t,x,p_{2})|\geq C |p_{1}-p_{2}|\,|t|.
\end{equation}
\end{Lemma} 
\noindent\textit{Proof: }The estimates (\ref{estimP}) and (\ref{estimsupp}) are proved for example in \cite{GSh1}, lemmas 1 and 2. 
The proof of (\ref{estimchar2}) is identical to the one of (\ref{estimchar1}) and the latter is given in lemma 5.6 of \cite{R}. \prfe

We will use repeatedly the following consequence of (\ref{estimP}). Assume that $\mathcal{P}(t)\leq\beta,\,\forall t\in\mathbb{R}$, for
some positive constant $\beta$. Then
\begin{equation}\label{suppchar}
|X(s,t,x,p)|\leq R+a(\beta)|s|,\quad\forall (x,p)\in\Xi(t),\,\forall (s,t)\in\mathbb{R}^{2},
\end{equation}
where 
\begin{equation}\label{a}
a(\beta)=\beta/\sqrt{1+\beta^{2}}<1.
\end{equation}
In fact by (\ref{charac1}), say for $s\geq 0$, $
|X(s)|\leq R+\sup_{0\leq \tau\leq s}|\widehat{P}(\tau,t,x,p)|\, s$. Moreover
\[
|\widehat{P}|^{2}=1-\frac{1}{1+|P|^{2}}\leq 1-\frac{1}{1+\beta^{2}}=\frac{\beta^{2}}{1+\beta^{2}}=a(\beta)^{2}
\]
and therefore $\sup_{0\leq \tau\leq s}|\widehat{P}(\tau,t,x,p)|\leq a(\beta)$. In particular, setting $s=t$ in (\ref{suppchar}),
\begin{equation}\label{suppdens}
f(t,x,p)=0, \textnormal{ for } |x|\geq R+a(\beta)|t|.
\end{equation}

\begin{Corollary}\label{estimatesdens}
Under the assumptions of lemma \ref{estimateschar},
\begin{equation}\label{derivpart}
\|\partial^j_{x}\partial^k_pf(t)\|_{\infty}\leq C\|\nabla\fzero\|_{\infty}(1+|t|)^k,\quad j+k=1, 
\end{equation}
\begin{equation}\label{decaysupp}
\textnormal{Vol}[\textnormal{supp}\, f(t,x,\cdot)]\leq C(1+|t|+|x|)^{-3}.
\end{equation}
\end{Corollary}
\noindent\textit{Proof: } 
The estimates (\ref{derivpart}) follow directly from (\ref{solvlasov}), (\ref{estimchar1}), (\ref{estimchar2}).  
For (\ref{decaysupp}) define
$\mathcal{A}(t,x)=\textnormal{supp}f(t,x,\cdot)=\{p\in \mathbb{R}^{3}:f(t,x,p) \neq 0 \}$.
By (\ref{solvlasov}) and (\ref{estimP}) we have 
\[
\mathcal{A}(t,x)\subseteq\{p:|p| \leq 2R\}\cap\{p:|X(0,t,x,p)|\leq R \}=\mathcal{U}\cap\mathcal{V}.
\]
For $|t|\leq 1$ we use that $\textrm{Vol}[\mathcal{A}(t,x)]\leq\textrm{Vol}[\mathcal{U}] \leq C \leq C(1+|t|)^{-3}$.
For $|t|>1$ we use that $\textrm{Vol}[\mathcal{A}(t,x)]\leq \textrm{Vol}\mathcal{[V]}$.
If $p_{1},p_{2} \in \mathcal{V}$ and $\eta\leq\eta_{0}$ then, by inequality (\ref{estimsupp}), 
\[
C|p_{1}-p_{2}|\,|t| \leq |X(0,t,x,p_{1})-X(0,t,x,p_{2})| \leq 2R.
\]
This means that the set $\mathcal{V}$ is contained in a ball with radius $C|t|^{-1}$, whose volume is then bounded 
by $C(1+|t|)^{-3}$. Moreover, for $|x|\leq R+|t|$ we have also $C(1+|t|)^{-3}\leq C(1+|t|+|x|)^{-3}$ and so (\ref{decaysupp}) is proved. 
\prfe

A key ingredient in the proof of theorem 1 is the analogue for the retarded solution of the integral representation formulae for the field
and the gradient of the field which have been proved in \cite{GS2}. We denote by $K_{\mathrm{ret}}$ the Lorentz force, 
$K_{\mathrm{ret}}=E_{\mathrm{ret}}+\widehat{p}\wedge B_{\mathrm{ret}}$.
\begin{Lemma}\label{repfield}
Assume $\mathcal{P}(t)\leq\beta$ for some positive constant $\beta$. 
Then there exist two smooth functions $\mathfrak{a}_{1}$,
$\mathfrak{a}_{2}$ uniformly bounded in the support of $f$ such that 
\begin{eqnarray}
E_{\mathrm{ret}}(t,x)=
&-&\int_{\Omega_{a}}\frac{dy}{|x-y|^{2}}\int_{|p|\leq\beta} dp\,\mathfrak{a}_{1}(\omega,p)f(t-|x-y|,y,p)\nonumber\\
&-&
\int_{\Omega_{a}}\frac{dy}{|x-y|}\int_{|p|\leq\beta} dp\,\mathfrak{a}_{2}(\omega,p)K_{\mathrm{ret}}f(t-|x-y|,y,p),\label{repfield0}
\end{eqnarray}
where $\omega=(y-x)/|y-x|$, $a=a(\beta)$ is given by (\ref{a}) and $\Omega_{a}$ denotes the set
\[
\Omega_{a}(t,x)=\{y\in\mathbb{R}^{3}:|y|\leq R+a|t-|x-y||\}.
\]
An analogous representation formula with two slightly different bounded kernels  
holds also for $B_{\mathrm{ret}}$.
\end{Lemma}
\noindent\textit{Sketch of the proof:} The proof of lemma \ref{repfield} is identical to the one of theorem 3 in \cite{GS2}, the kernels of
the integral representations being also the same. 
We give here the idea of the proof for sake of completeness. 
By (\ref{eletfield}) and (\ref{sources}) we have
\begin{equation}\label{repfield1}
E^{i}_{\mathrm{ret}}(t,x)=-\int_{\Omega_{a}} \frac{dy}{|x-y|}\int_{|p|\leq\beta} dp\,(\partial_{y_{i}}f+\widehat{p}_{i}\partial_{t}f)(t-|x-y|,y,p).
\end{equation}
To justify that the integral w.r.t. $y$ in (\ref{repfield1}) is extended over the set $\Omega_{a}$ we notice that, by (\ref{suppdens}), 
$f(t-|x-y|,y,p)=0$ for $|y|\geq R+a|t-|x-y||$. In particular, since $a<1$, then $\Omega_a(t,x)$ is bounded for any fixed $t\in\mathbb{R}$ and
$x\in\mathbb{R}^3$.
Now we express $\partial_{y_i}f(t-|x-y|,y,p)$ and $\partial_{t}f(t-|x-y|,y,p)$ in terms of the perfect derivatives 
of $f(t-|x-y|,y,p)$ via the identities
\begin{eqnarray}
\partial_{y_{i}}f(t-|x-y|,y,p)&=&\omega_{i}(1+\omega\cdot\widehat{p})^{-1}Tf(t-|x-y|,y,p)\nonumber\\
&&+
\Bigg(\delta_{ik}-\frac{\omega_{i}\widehat{p}_{k}}{1+\omega\cdot\widehat{p}}\Bigg) \partial_{y_{k}}[f(t-|x-y|,y,p)],\label{identity1}
\end{eqnarray}
\begin{eqnarray}
\partial_{t}f(t-|x-y|,y,p)&=&(1+\omega\cdot\widehat{p})^{-1}Tf(t-|x-y|,y,p)\nonumber\\
&&-
\frac{\widehat{p}_{k}}{1+\omega\cdot\widehat{p}} \partial_{y_{k}}[f(t-|x-y|,y,p)].\label{identity2}
\end{eqnarray}
We substitute (\ref{identity1}) and (\ref{identity2}) into (\ref{repfield1}) and integrate by parts. 
Since $f$ vanishes on the boundary of $\Omega_{a}$, we obtain  
\begin{eqnarray}
E_{\mathrm{ret}}(t,x)=
&-&\int_{\Omega_{a}}\frac{dy}{|x-y|^{2}}\int_{|p|\leq\beta} dp\,\mathfrak{a}_{1}(\omega,p)f(t-|x-y|,y,p)\nonumber\\
&-&
\int_{\Omega_{a}}\frac{dy}{|x-y|}\int_{|p|\leq\beta} dp\,\mathfrak{b}(\omega,p)Tf(t-|x-y|,y,p),\label{repfield2}
\end{eqnarray}
where
\begin{eqnarray}
&&\mathfrak{a}_{1}(\omega,p)=\frac{\omega+\widehat{p}}{(1+p^2)
(1+\omega\cdot\widehat{p})^2},\label{kernel1}\\
&&\mathfrak{b}(\omega,p)=\frac{\omega+\widehat{p}}{1+\omega\cdot\widehat{p}}\,.\label{kernel2}
\end{eqnarray}
By (\ref{vlasov2}), $Tf=-K_{\mathrm{ret}}\cdot\nabla_{p}f$. Substituting into (\ref{repfield2}) and integrating by parts in $p$, we get 
(\ref{repfield0}) with
$\mathfrak{a}_{2}=\partial_{p}\mathfrak{b}$ (again, since $f$ vanishes for $|p|=\beta$ there are no boundary terms). 
The kernels $\mathfrak{a}_{1}$, $\mathfrak{a}_{2}$ are bounded by $C\sqrt{1+p^{2}}$ (see \cite{GS3}). Thus in the present case, 
because of our assumption $\mathcal{P}(t)\leq \beta$, they are uniformly bounded. \prfe

The following lemma contains the analogous representation for the derivatives of the retarded field and corresponds to theorem 4 of \cite{GS2}:
\begin{Lemma}\label{repderfield}
Assume $\mathcal{P}(t)\leq\beta$ for some positive constant $\beta$. Then there exist smooth functions $\mathfrak{b}_{1}$, 
$\mathfrak{b}_{2}$, $\mathfrak{b}_{3}$ uniformly bounded in the support of $f$ such that
\begin{eqnarray}
DE_{\mathrm{ret}}(t,x)=
&-&\int_{\Omega_{a}}\frac{dy}{|x-y|^{3}}\int_{|p|\leq\beta} dp\,\mathfrak{b}_{1}(\omega,p)f(t-|x-y|,y,p)\nonumber\\
&-&\int_{\Omega_{a}}\frac{dy}{|x-y|^{2}}\int_{|p|\leq\beta} dp\,\mathfrak{b}_{2}(\omega,p)K_{\mathrm{ret}}f(t-|x-y|,y,p)\label{repderfield0}\\
&-&\int_{\Omega_{a}}\frac{dy}{|x-y|}\int_{|p|\leq\beta} dp\,\mathfrak{b}_{3}(\omega,p)D(K_{\mathrm{ret}}f)(t-|x-y|,y,p).\nonumber
\end{eqnarray}
Moreover the kernel $\mathfrak{b}_{1}(\omega,p)$ satisfies
\begin{equation}\label{zeroaverage}
\int_{S^{2}}\mathfrak{b}_{1}(\omega,p)d\omega=0.
\end{equation}
The derivatives of $B_{\mathrm{ret}}$ admit a similar representation with three different bounded kernels $\mathfrak{b}_{1}'$, $\mathfrak{b}_{2}'$,
$\mathfrak{b}_{3}'$ and $\mathfrak{b}_{1}'$ also satisfies (\ref{zeroaverage}).  
\end{Lemma}  
\noindent\textit{Sketch of the proof:} Let $I_{1}$, $I_{2}$ denote the two integrals in (\ref{repfield0}). 
By differentiating $I_{2}$ we obtain the
third term in (\ref{repderfield0}) with $\mathfrak{b}_{3}=\mathfrak{a}_{2}$.
Differentiating $I_{1}$ we get
\[
DI_{1}(t,x)=-\int_{\Omega_{a}}\frac{dy}{|x-y|^{2}}\int_{|p|\leq\beta} dp\,\mathfrak{a}_{1}(\omega,p)Df(t-|x-y|,y,p).
\]
The absence of boundary terms is again due to the fact that $f$ vanishes on the boundary of $\Omega_{a}$.
In the previous expression we use again (\ref{identity1}), (\ref{identity2}) and then integrate by parts. We end up with 
(\ref{repderfield0}) after defining properly the 
various kernels. The exact form of the latter quantities is given in \cite{GS2} but here it is not important; the crucial point is that the 
kernels are uniformly bounded for
$|p|\leq\beta$. The identity (\ref{zeroaverage}) is proved in \cite{GS2}. \prfe

\section{Estimates on the retarded field and uniqueness}\label{estimates}
\setcounter{equation}{0}
All the estimates in this paper will be based on the following two lemmas:
\begin{Lemma}\label{basiclemma1}
Let $I_{n}^{q}(t,x)$ ($n=1,2$; $q\in\mathbb{R}$) denote the integral
\[
I_{n}^{q}(t,x)=\int\frac{dy}{|x-y|^{n}}(1+|t-|x-y||+|y|)^{-q}.
\]
Then for all $(t,x)\in\mathbb{R}_{t}\times\mathbb{R}^{3}_{x}$ the following estimates hold:
\begin{eqnarray*}
&&I_{1}^{q}\leq C(1+|t|+|x|)^{-1}(1+|t-|x||)^{-q+3},\quad q>3,\\
&&I_{2}^{q}\leq C(1+|t|+|x|)^{-1}(1+|t-|x||)^{-q+2},\quad q>2.
\end{eqnarray*}
\end{Lemma}

\begin{Lemma}\label{basiclemma2}
Assume $g\in C^1\cap L^\infty(\mathbb{R}_t\times\mathbb{R}^3_x\times\mathbb{R}^3_p)$ and vanishes for $|p|\geq\beta$.
Assume also that  
\begin{equation}\label{hyp1}
\textnormal{Vol}[\textnormal{supp}\, g(t,x,\cdot)]\leq C(1+|t|+|x|)^{-3},
\end{equation}
\begin{equation}\label{hyp2}
Dg\in L^{\infty}(\mathbb{R}_{t}\times\mathbb{R}^{3}_{x}\times\mathbb{R}^{3}_{p}).
\end{equation}
Let $\mathfrak{b}_{1}(\omega,p)$ be smooth and satisfy (\ref{zeroaverage}).
Then the integral
\begin{equation}\label{singintegral}
I(t,x)=\int\frac{dy}{|x-y|^{3}}\int_{|p|\leq\beta} dp\,\mathfrak{b}_{1}(\omega,p)g(t-|x-y|,y,p),
\end{equation}
satisfies the estimate
\[
|I(t,x)|\leq C_{*} (\|g\|_{\infty}+\|Dg\|_{\infty})(1+|t|+|x|)^{-1}(1+|t-|x||)^{-7/4},
\]
for all $(t,x)\in\mathbb{R}_{t}\times\mathbb{R}^{3}_{x}$, where $C_{*}=C_{*}(\beta)$.
\end{Lemma}

The quite long and technical proofs of lemmas \ref{basiclemma1} and \ref{basiclemma2} are postponed in appendix. 

We denote by $\|F\|_{w}$ the weighted norm:
\begin{eqnarray}
\|F\|_{w}=\sup_{t,x}[(1+|t|+|x|)(1+|t-|x||)^w|F(t,x)|],\label{norm}
\end{eqnarray}
where $w>0$ and set $F_{\mathrm{ret}}=(E_{\mathrm{ret}},B_{\mathrm{ret}})$. In the following two propositions we estimate the retarded field
generated by a solution $f_{\mathrm{ret}}$ of RVM$_{\mathrm{ret}}$ with initial data and regularity as stated in theorem \ref{main}.
\begin{Proposition}\label{estimatefield}
Assume $\mathcal{P}(t)\leq\beta$ and (\ref{hyp1}) holds for $g\equiv f_{\mathrm{ret}}$. 
Then there exists a constant $\varepsilon>0$ which depends on $R$ and $\beta$ such that for
$\|\fzero\|_{\infty}\leq\varepsilon$ the retarded field satisfies the estimate
\begin{equation}\label{estimretfield}
\|F_{\mathrm{ret}}\|_{1}\leq C_{*}\|\fzero\|_{\infty},
\end{equation}
where $C_{*}=C_{*}(R,\beta)$.
\end{Proposition}
\noindent\textit{Proof: }
Using (\ref{hyp1}) to estimate (\ref{repfield0}) we get, with the notation of lemma \ref{basiclemma1},
\[
|E_{\mathrm{ret}}(t,x)|\leq C_{*} \|\fzero\|_{\infty}I_{2}^{3}(t,x)
+C_{*} \|\fzero\|_{\infty}\int_{\Omega_{a}}\frac{dy}{|x-y|}
\frac{|F_{\mathrm{ret}}(t-|x-y|,y)|}{(1+|t-|x-y||+|y|)^{3}}.
\]
An analogous estimate holds for $B_{\mathrm{ret}}$ and so we have
\begin{eqnarray*}
|F_{\mathrm{ret}}(t,x)|&\leq& C_{*}\|\fzero\|_{\infty} I_{2}^{3}(t,x)+C_{*} \|\fzero\|_{\infty}\int_{\Omega_{a}}\frac{dy}{|x-y|}
\frac{|F_{\mathrm{ret}}(t-|x-y|,y)|}{(1+|t-|x-y||+|y|)^{3}}\\
&\leq& C_{*}\|\fzero\|_{\infty}I_{2}^{3}(t,x)+ C_{*} \|\fzero\|_{\infty}\|F_{\mathrm{ret}}\|_1\\
&&\times\int_{\Omega_{a}}\frac{dy}{|x-y|}
(1+|t-|x-y||+|y|)^{-4}(1+|t-|x-y|-|y||)^{-1}\\
&\leq& C_{*}\|\fzero\|_{\infty}I_{2}^{3}(t,x)+C_{*}\|\fzero\|_{\infty}\|F_{\mathrm{ret}}\|_1 I_{1}^{5}(t,x).
\end{eqnarray*}
Here we used that 
\begin{equation}\label{trick}
(1+|t-|x-y|-|y||)^{-1}\leq C_{*}(1+|t-|x-y||+|y|)^{-1}
\end{equation}
holds for $y\in\Omega_{a}$. In fact
\begin{eqnarray*}
\frac{1+|t-|x-y||+|y|}{1+|t-|x-y|-|y||}&\leq& (1+2R)\frac{1+|t-|x-y||+|y|}{1+2R+|t-|x-y|-|y||}\\
&\leq& (1+2R)\frac{1+R+(1+a)|t-|x-y||}{1+R+(1-a)|t-|x-y||}\\
&\leq& 2\bigg (\frac{1+2R}{1-a}\bigg )=C_{*}.
\end{eqnarray*}
Hence, by lemma \ref{basiclemma1}
\[
(1+|t|+|x|)(1+|t-|x||)|F_{\mathrm{ret}}(t,x)|\leq C_{*}\|\fzero\|_{\infty}+C_{*}\|\fzero\|_{\infty}\|F_{\mathrm{ret}}\|_1
\]
and so
$(1-C_{*}\|\fzero\|_{\infty})\|F_{\mathrm{ret}}\|_1\leq C_{*}\|\fzero\|_{\infty}$, 
which implies (\ref{estimretfield}) for $\|\fzero\|_{\infty}\leq 1/2C_{*}$. \prfe

By the same argument we can prove the following \textit{a priori} estimate on the derivatives of the field.
\begin{Proposition}\label{estimatederfield}
Assume $\mathcal{P}(t)\leq\beta$ and (\ref{hyp1}), (\ref{hyp2}) hold for $g\equiv f_{\mathrm{ret}}$.
Then for a proper small $\|\fzero\|_{\infty}$, the retarded field satisfies
\begin{equation}\label{estimderretfield}
\|DF_{\mathrm{ret}}\|_{7/4}\leq C_{*}z,
\end{equation}
where $z=(1+\|\fzero\|_{\infty})(\|\fzero\|_{\infty}+\|Df\|_{\infty})$ and $C_{*}=C_{*}(R,\beta)$.
\end{Proposition}
\noindent\textit{Proof: } By (\ref{repderfield0}) we have,
\begin{eqnarray*}
|DE_{\mathrm{ret}}(t,x)|&\leq& |I|+|II|+|III|\\
&&+
C_{*}\|f\|_{\infty}\int_{\Omega_{a}}\frac{dy}{|x-y|}\frac{|DF_{\mathrm{ret}}(t-|x-y|,y)|}{(1+|t-|x-y||+|y|)^{3}},
\end{eqnarray*} 
where $I$ is the integral (\ref{singintegral}) with $g\equiv f_{\mathrm{ret}}$  and 
\begin{eqnarray*}
II(t,x)&=&\int_{\Omega_{a}}\frac{dy}{|x-y|^{2}}\int_{|p|\leq\beta}dp\,\mathfrak{b}_{2}(\omega,p)K_{\mathrm{ret}}f(t-|x-y|,y,p),\\
III(t,x)&=&\int_{\Omega_{a}}\frac{dy}{|x-y|}\int_{|p|\leq\beta} dp\,\mathfrak{b}_{3}(\omega,p)K_{\mathrm{ret}}Df(t-|x-y|,y,p).
\end{eqnarray*}
To estimate $II$ and $III$  we use (\ref{estimretfield}) and (\ref{trick}).
So doing we get
\[
II(t,x)\leq C_{*}\|\fzero\|_{\infty}^{2}I_{2}^{5}(t,x),\quad III(t,x)\leq C_{*}\|\fzero\|_{\infty}\|Df\|_{\infty}I_{1}^{5}(t,x)
\]
and therefore, using lemmas \ref{basiclemma1} and \ref{basiclemma2},
\begin{eqnarray*}
|DF_{\mathrm{ret}}(t,x)|&\leq&C_{*}z(1+|t|+|x|)^{-1}(1+|t-|x||)^{-7/4}\\
&&+C_{*}\|\fzero\|_{\infty}\int_{\Omega_a}\frac{dy}{|x-y|}\frac{|DF_{\mathrm{ret}}(t-|x-y|,y)|}{(1+|t-|x-y||+|y|)^{3}}\\
&\leq& \frac{C_{*}z}{(1+|t|+|x|)(1+|t-|x||)^{7/4}}+C_{*}\|\fzero\|_{\infty}\|DF_{\mathrm{ret}}\|_{7/4}I_{1}^{23/4}.
\end{eqnarray*}
Hence, by lemma \ref{basiclemma1},  
\[
\|DF_{\mathrm{ret}}\|_{7/4}(1-C_{*}\|\fzero\|_{\infty})\leq C_{*}z,
\]
which concludes the proof. \prfe

We also notice that (\ref{estimretfield}), (\ref{estimderretfield}) implies FSC w.r.t. $\eta=C_{*}z$. In particular for the
approximation sequence defined in section \ref{existence} below we will have $\eta=C\Delta$ for a proper small $\Delta$.

To conclude this section we prove the uniqueness part of theorem 1:
\begin{Proposition}\label{unique}
Let $\fzero(x,p)\geq 0$ be given in $C_{0}^{1}(\mathbb{R}^{3}_{x}\times\mathbb{R}^{3}_{p})$ and consider the following class of 
solutions of RVM:
\begin{eqnarray*}
\mathfrak{D}(\fzero, \eta)=\big\{ (f,F)\in C^{1}:\, &&f(0,x,p)=\fzero(x,p),\\
&&\|\fzero\|_{\infty}+\|\nabla\fzero\|_{\infty}\leq\eta,\\
&&(f,F)\textnormal{ satisfies FSC w.r.t }\eta,\\
&&F(t,\cdot)\in L^{2}(\mathbb{R}^{3}),\, \|F(t,\cdot)\|_{L^2}\to 0\textnormal{ as }t\to -\infty\}.
\end{eqnarray*}
Then there exists a positive constant $\eta_{0}$ such that for $\eta\leq\eta_{0}$ either $\mathfrak{D}(\fzero,\eta)$ is empty or it 
contains only one element. 
\end{Proposition}
\noindent\textit{Proof: }Let $(f_{1},E_{1},B_{1})$ and $(f_{2},E_{2},B_{2})$ be two solutions of RVM in $\mathfrak{D}(\fzero,\eta)$ 
and put $\delta f=(f_{1}-f_{2}),\,\delta E=(E_{1}-E_{2}),\,\delta B=(B_{1}-B_{2})$.
Then $(\delta f,\delta E, \delta B)$ satisfies the system
\begin{equation}\label{deltavlasov}
\partial_{t}\delta f+\widehat{p}\cdot\partial_{x}\delta f+(E_{1}+\widehat{p}\wedge B_{1})\cdot\partial_{p}\delta f=-(\delta
E+\widehat{p}\wedge\delta B)\cdot\partial_{p} f_{2},
\end{equation}
\begin{equation}\label{deltamaxwell}
\left\{\begin{array}{ll}\partial_{t}\delta E=\partial_{x}\wedge\delta B-4\pi\delta j,& \quad \partial_{x}\cdot\delta E=4\pi\delta\rho,\\
\partial_{t}\delta B=-\partial_{x}\wedge\delta E,& \quad\partial_{x}\cdot\delta B=0,\end{array}\right.
\end{equation}
with initial data $\delta f(0,x,p)=0$ and where $\delta\rho=\int dp\,\delta f,\, \delta j=\int dp\,\widehat{p}\,\delta f$.
Our aim is to show that $\delta f=\delta E=\delta B\equiv 0$. However we remark at this point  that it is sufficient to prove this for  
$t\leq 0$. For, if the uniqueness holds in the past, then $(f_{i},E_{i},B_{i})$, $i=1,2$, will be solutions of RVM with the same initial 
data and then, since for a proper small $\eta$ the estimate $\mathcal{P}(t)\leq 2R $ is satisfied for all $t\geq 0$ (see lemma
\ref{estimateschar}),  
the uniqueness in the future follows by \cite{GS2}. 
Hence we assume $t\leq 0$ in the rest of the proof. 
The $L^{2}$ solution $\delta F=(\delta E,\delta B)$ of (\ref{deltamaxwell}) which 
satisfies $\|\delta F(t,\cdot)\|_{L^2}\to 0$ for $t\to -\infty$ is unique, because the $L^{2}$ norm of a solution of the \textit{homogeneous
} Maxwell equations is constant. We claim that, for a proper small $\eta$, this solution is represented by
\begin{eqnarray}\label{deltafield}
\delta E(t,x)&=&-\int\frac{dy}{|x-y|}(\partial_{x}\delta\rho+\partial_{t}\delta j)(t-|x-y|,y),\label{deltafield1}\\
\delta B(t,x)&=&\int\frac{dy}{|x-y|}(\partial_{x}\wedge\delta j)(t-|x-y|,y).\label{deltafield2}
\end{eqnarray} 
(Note that (\ref{deltafield1}), (\ref{deltafield2}) define a solution of (\ref{deltamaxwell}) because $\delta\rho,\, \delta j$ satisfy the continuity equation 
$\partial_{t}\delta\rho +\partial_{x}\cdot\delta j=0$ as a consequence of (\ref{deltavlasov})). To this purpose we first note that 
$\|\delta f(t)\|_{\infty}\leq 2\|\fzero\|_{\infty}$ and that, for a proper small $\eta$,
\begin{eqnarray}
&&\delta f(t,x,p)=0,\quad\textnormal{ for }|x|\geq R-a(2R)t,\\
&&\textnormal{Vol}[\textnormal{supp}\, \delta f(t,x,\cdot)]\leq C(1-t+|x|)^{-3},\label{decaysuppdelta}\\
&&\|\partial_x^j\partial_{p}^k\delta f(t)\|_{\infty}\leq C\|\nabla\fzero\|_{\infty}(1-t)^k,\quad j+k=1,\label{derdelta}
\end{eqnarray} 
cf. (\ref{suppdens}) and corollary \ref{estimatesdens}.
Moreover, the function (\ref{deltafield}) admits an integral representation formula similar to that one given in lemma \ref{repfield}:
\begin{eqnarray}
\delta E(t,x)&=&-\int\frac{dy}{|x-y|^{2}}\int dp\,\mathfrak{a}_{1}(\omega,p)\delta f(t-|x-y|,y,p)\nonumber\\
&&-\int\frac{dy}{|x-y|}\int dp\,\mathfrak{b}(\omega,p)T\delta f(t-|x-y|,y,p),\label{repdeltafield}
\end{eqnarray}
cf. (\ref{repfield2}) (it is understood that the integrals in $y$ are over $\Omega_{a}$ and the ones in $p$ are over $\{|p|\leq 2R\}$).
By (\ref{deltavlasov}) we have
\[
T\delta f=-(E_{1}+\widehat{p}\wedge B_{1})\cdot\partial_{p}\delta f-(\delta E+\widehat{p}\wedge\delta B)\cdot\partial_{p} f_{2}.
\]
Substituting into (\ref{repdeltafield}) and integrating by parts in $p$ we get
\begin{eqnarray}
\delta E(t,x)&=&-\int\frac{dy}{|x-y|^{2}}\int dp\,\mathfrak{a}_{1}(\omega,p)\delta f(t-|x-y|,y,p)\nonumber\\
&&-\int\frac{dy}{|x-y|}\int dp\, \mathfrak{a}_{2}(\omega,p)(E_{1}+\widehat{p}\wedge B_{1})\delta f(t-|x-y|,y,p)\nonumber\\
&&-\int\frac{dy}{|x-y|}\int dp\, \mathfrak{a}_{2}(\omega,p)(\delta E+\widehat{p}\wedge \delta B) f_{2}(t-|x-y|,y,p)\nonumber\\
&=&I+II+III,\label{repdeltafield2}
\end{eqnarray}
where $\mathfrak{a}_{2}=\partial_{p}\mathfrak{b}$ (cf. (\ref{kernel2})). By (\ref{decaysuppdelta}), the integral $I$ is bounded by
\[
|I(t,x)|\leq C\|\delta f\|_{\infty}I_{2}^{3}(t,x)\leq C\|\fzero\|_{\infty}(1-t+|x|)^{-2}.
\]
To estimate $II(t,x)$ we use that for $y\in\Omega_{a}$ the free streaming condition in the past gives
\begin{eqnarray*}
|(E_{1}+\widehat{p}\wedge B_{1})|&\leq&\eta(1-t+|x-y|+|y|)^{-\alpha}(1+R-t+|x-y|-|y|)^{-\alpha}\\
&\leq& C(1-t+|x-y|+|y|)^{-2\alpha}\leq C (1-t+|x-y|+|y|)^{-1}, 
\end{eqnarray*} 
since $\alpha>\frac{1}{2}$. The same applies for $\delta E+\widehat{p}\wedge \delta B$ in $III(t,x)$ and so we get
\begin{displaymath}
|II(t,x)|+|III(t,x)|\leq C(\|\delta f\|_{\infty}+\|\fzero\|_{\infty})I_{1}^{4}(t,x)\leq C\|\fzero\|_{\infty}(1-t+|x|)^{-2}.
\end{displaymath}
Substituting these estimates into (\ref{repdeltafield2}) and using the same argument for $\delta B$ we get
\[
|\delta F|\leq C\|\fzero\|_{\infty}(1-t+|x|)^{-2}
\]
and so $\|\delta F(t,\cdot)\|_{L^2}\to 0$ as $t\to -\infty$, as we claimed. We are able now to complete the proof of proposition \ref{unique}. 
Let us introduce 
\[
\|\delta F\|'=\sup_{x;t\leq 0}[(1-t+|x|)|\delta F(t,x)|].
\]
By (\ref{repdeltafield2}) we have
\begin{equation}\label{estdeltafield}
|\delta E(t,x)|\leq C(\|\delta f\|_{\infty}+\|\fzero\|_{\infty}\|\delta F\|') (1-t+|x|)^{-2}.
\end{equation}
On the other hand, integrating (\ref{deltavlasov}) along the characteristics of the Vlasov equation and using (\ref{derdelta}) we get
\begin{equation}\label{estdeltadens}
\|\delta f\|_{\infty}\leq C\|\nabla\fzero\|_{\infty}\|\delta F\|'(1-t).
\end{equation}
Combining (\ref{estdeltafield}) and (\ref{estdeltadens}) we get
\[
|\delta E(t,x)|\leq C[\|\nabla\fzero\|_{\infty}+\|\fzero\|_{\infty}]\|\delta F\|'(1-t+|x|)^{-1}.
\]
Hence, from the analogous estimate on $\delta B$ we find
\[
\|\delta F\|'\leq C\eta\|\delta F\|_{0}'
\]
which entails $\|\delta F\|'=0$ for $\eta<C^{-1}$ and thus $\delta F=\delta f=0$. \prfe

We remark that the meaning of the last condition in the definition of $\mathfrak{D}(\fzero, \eta)$ is that all the energy is contained in the
particles in the limit $t\to -\infty$.
However this energy is not carried to past null infinity since the particles always move, even asymptotically, at velocities strictly 
smaller than the speed of light.
The solution of theorem \ref{main} belongs to the class $\mathfrak{D}(\fzero,C\Delta)$ and therefore, for a proper small $\Delta$, it
is unique in this class.

\section{Proof of existence}\label{existence}
\setcounter{equation}{0}
The existence part of theorem \ref{main} is proved by a standard recursive argument which we split in three steps:
\subsection*{Step 1: The approximation sequence}
We define: $f_{1}(t,x,p)=\fzero(x-\widehat{p}\,t,p)$,
\begin{eqnarray*}
E_{1}(t,x)&=&\int \frac{dy}{|x-y|}[-\partial_{x}\rho_{1}-\partial_{t}j_{1}](t-|x-y|,y),\\
B_{1}(t,x)&=&\int \frac{dy}{|x-y|}[\partial_{x}\wedge j_{1}](t-|x-y|,y),
\end{eqnarray*}
where $\rho_{1}=\int dp f_{1},\, j_{1}=\int dp \,\widehat{p}f_{1}$ and set $F_{1}=(E_{1},B_{1})$. 
This solution corresponds to the case in which the particles do not
interact with the field, i.e. the force term in (\ref{vlasov2}) is omitted.
Now, supposing that $f_{n}$ is been defined, we build $\rho_{n},j_{n},E_{n},B_{n}$ via the formulae
$\rho_{n}=\int dpf_{n}$, $j_{n}=\int dp\,\widehat{p}f_{n}$,
\begin{eqnarray*}
E_{n}(t,x)&=&\int \frac{dy}{|x-y|}[-\partial_{x}\rho_{n}-\partial_{t}j_{n}](t-|x-y|,y),\\
B_{n}(t,x)&=&\int \frac{dy}{|x-y|}[\partial_{x}\wedge j_{n}](t-|x-y|,y)
\end{eqnarray*}
and put $F_{n}=(E_{n},B_{n})$.
Now consider the following initial value problem for the function $(X,P):\mathbb{R}_{s}\to \mathbb{R}^{6}$:
\begin{eqnarray}
&&\frac{dX}{ds}=\widehat{P},\quad\frac{dP}{ds}=E_{n}+\widehat{P}\wedge B_{n},\label{charn}\\
&&(X(t),P(t))=(x,p).
\end{eqnarray}
Let $(X_{n+1}(s,t,x,p),P_{n+1}(s,t,x,p))$ denote the classical solution of the previous problem
(sometimes it will be denoted by $(X_{n+1}(s),P_{n+1}(s))$ for short) and define $f_{n+1}$ as
\begin{equation}\label{densn}
f_{n+1}(t,x,p)=\fzero(X_{n+1}(0,t,x,p),P_{n+1}(0,t,x,p)).
\end{equation}
$f_{n+1}$ solves the following \emph{linear} equation:
\begin{equation}\label{vlasovn}
\partial_{t}f_{n+1}+\widehat{p}\cdot\partial_{x}f_{n+1}+(E_{n}+\widehat{p}\wedge
B_{n})\cdot\partial_{p}f_{n+1}=0,
\end{equation}
with initial datum $f_{n+1}(0,x,p)=\fzero(x,p)$.
The following lemma is easily proved by induction:
\begin{Lemma}\label{approx}
For a proper small $\Delta$, the sequence $(f_{n},F_{n})$ is constituted by $C^{2}$ functions and the following estimates hold $\forall\,
n\in\mathbb{N}$:
For $j,k\in\{0,1,2\}$, $0\leq j+k\leq 2$, 
\begin{equation}\label{estimdensn}
\|D^{j}\partial_{p}^{k}f_{n}(t)\|_{\infty}\leq C\Delta (1+|t|)^{k},\quad\forall\, t\in\mathbb{R};
\end{equation}
for all $(t,x)\in\mathbb{R}_{t}\times\mathbb{R}^{3}_{x}$:
\begin{equation}\label{estimPn}
\mathcal{P}_{n}(t)=\sup_{0\leq s\leq t}\{|p|:f_{n}(s,x,p)\neq 0,\textnormal{ for some }x\}\leq 2R,
\end{equation}
\begin{equation}\label{decaysuppn}
\textnormal{Vol}[\textnormal{supp}\, f_{n}(t,x,\cdot)]\leq C(1+|t|+|x|)^{-3},
\end{equation}
\begin{equation}\label{estimate1n}
|F_{n}(t,x)|\leq C\Delta (1+|t|+|x|)^{-1}(1+|t-|x||)^{-1},
\end{equation}
\begin{equation}
|D^{k}F_{n}(t,x)|\leq C\Delta (1+|t|+|x|)^{-1}(1+|t-|x||)^{-7/4}, \quad k=1,2.\label{estimate2n}
\end{equation}
\end{Lemma}

\noindent\textit{Proof: }The estimates (\ref{estimdensn}), (\ref{estimPn}) and (\ref{decaysuppn}) in the case $n=1$ follow directly from the 
definition of $f_{1}$. 
For (\ref{estimate1n}) in the case $n=1$, note that the integral representation formula for $E_{1}$ reduces to the first integral of
(\ref{repfield0}), namely
\[
E_{1}(t,x)=-\int\frac{dy}{|x-y|^{2}}\int dp\,\mathfrak{a}_{1}(\omega,p)f_{1}(t-|x-y|,y,p).
\]
(From now on it will be understood that the integrals in $p$ are over the set $\{|p|\leq 2R\}$ and the ones in $y$ over
$\Omega_{a}(t,x)$, with $a=a(2R)$). 
The previous integral is bounded by $C\Delta I_{2}^{3}(t,x)$, i.e., using lemma 4,
\[
|E_{1}(t,x)|\leq C\Delta (1+|t|+|x|)^{-1}(1+|t-|x||)^{-1}.
\]
The same is true for $B_{1}$ and therefore (\ref{estimate1n}) in the case $n=1$ is proved. Analogously, the representation formula for 
$DE_{1}$
reduces to the integral (\ref{singintegral}), with $g\equiv f_{1}$ and therefore lemma \ref{basiclemma2} gives (\ref{estimate2n})$_{k=1}$ in 
the case $n=1$. In a similar way, for $D^{2}E_{1}$ one has 
\[
D^{2}E_{1}(t,x)=\int\frac{dy}{|x-y|^{3}}\int dp\,\mathfrak{b}_{1}(\omega,p)Df_{1}(t-|x-y|,y,p),
\]
which, applying lemma \ref{basiclemma2} to $g\equiv Df_1$, is estimated by
\begin{eqnarray*}
|D^{2}E_{1}(t,x)|&\leq& C\frac{(\|Df_{1}\|_{\infty}+\|D^{2}f_{1}\|_{\infty})}{(1+|t|+|x|)(1+|t-|x||)^{7/4}}\\
&\leq& C\Delta (1+|t|+|x|)^{-1}(1+|t-|x||)^{-7/4}.
\end{eqnarray*}
Now assume that (\ref{estimdensn})--(\ref{estimate2n}) hold for $(f_{n},F_{n})$. Since FSC is satisfied for $\eta=C\Delta$, 
then for a proper small $\Delta$ the
estimates (\ref{estimPn}), (\ref{decaysuppn}) follow by lemma \ref{estimateschar} and corollary \ref{estimatesdens}. The same is true for 
(\ref{estimdensn}) in the case $j+k\leq 1$, while the case $j+k=2$ follows by
using the estimates on the second derivatives of $F_{n}$. Precisely, the argument for the proof of lemma \ref{estimateschar} 
shows that the characteristics satisfy the estimate
\[
|\partial_{x}^{j}\partial_{p}^{k}(X,P)(0,t,x,p)|\leq C(1+|t|)^{k},\quad\textnormal{for }j+k=2,
\]
provided that the field satisfies FSC (for a proper small $\eta$) and
\[
|\partial^{2}_{x}F(t,x)|\leq C (1+|t|+|x|)^{-\alpha} (1+|t|-|x|)^{-\alpha-1}.
\]
The details are omitted because they are the same as for the proof of lemma \ref{estimateschar}.
To complete the proof of (\ref{estimate1n}) we apply lemma \ref{basiclemma1} to the representation formula
\begin{eqnarray*}
E_{n+1}(t,x)&=&-\int\frac{dy}{|x-y|^{2}}\int dp\,\mathfrak{a}_{1}(\omega,p)f_{n+1}(t-|x-y|,y,p)\\
&&-\int\frac{dy}{|x-y|}\int dp\,\mathfrak{a}_{2}(\omega,p)(E_{n}+\widehat{p}\wedge B_{n})f_{n+1}(t-|x-y|,y,p),
\end{eqnarray*}
which is proved as lemma \ref{repfield}. Similar equations can be written for the first and the second derivatives of $E_{n+1}$. 
By applying lemmas \ref{basiclemma1} and \ref{basiclemma2} to these equations, 
the estimate (\ref{estimate2n}) follows after a straightforward argument. \prfe 

\subsection*{Step 2: Convergence in the $C^{0}$ norm}
In this step we will prove the convergence of the sequence $F_{n}$ with respect to the norm $\|\,\cdot\,\|_{3/4}$. 
Indeed the convergence holds in the norm (\ref{norm}) for all $0<w<1$. The choice $w=3/4$ suffices for our purpose and it is made only 
for sake of simplicity.
\begin{Proposition}\label{convergence0}
For properly small initial data, the sequence $F_{n}$ converges in the norm $\|\,\cdot\,\|_{3/4}$.
\end{Proposition}
\noindent\textit{Proof:} Put $\delta f_{n,m}=f_{n}-f_{m}$ and $\delta F_{n,m}=F_{n}-F_{m}$.
The analogue of (\ref{repfield0}) for the approximation sequence is
\begin{eqnarray}
E_{n}(t,x)&=&-\int\frac{dy}{|x-y|^{2}}\int dp\,\mathfrak{a}_{1}(\omega,p)f_{n}(t-|x-y|,y,p)\nonumber\\
&&-\int\frac{dy}{|x-y|}\int dp\,\mathfrak{a}_{2}(\omega,p)K_{n-1}f_{n}(t-|x-y|,y,p),\label{repfield0n}
\end{eqnarray}
where $K_{n-1}=E_{n-1}+\widehat{p}\wedge B_{n-1}$. Thus
\begin{eqnarray*}
\delta E_{n,m}&=&-\int\frac{dy}{|x-y|^{2}}\int dp\,\mathfrak{a}_{1}(\omega,p)\delta f_{n,m}(t-|x-y|,y,p)\\
&&-\int\frac{dy}{|x-y|}\int dp\,\mathfrak{a}_{2}(\omega,p)K_{n-1}\delta f_{n,m}(t-|x-y|,y,p)\\
&&-\int\frac{dy}{|x-y|}\int dp\,\mathfrak{a}_{2}(\omega,p)\delta K_{n-1,m-1}f_{m}(t-|x-y|,y,p).
\end{eqnarray*}
Estimating:
\begin{eqnarray*}
|\delta E_{n,m}|&\leq& C \Bigg (\int \frac{dy}{|x-y|^{2}}\int dp\, |\delta f_{n,m}|(t-|x-y|,y,p) \\
&&+ \int  \frac{dy}{|x-y|}\int dp\, |F_{n-1}|\,|\delta f_{n,m}|(t-|x-y|,y,p)\\
&&+\int \frac{dy}{|x-y|}\int dp\, |\delta F_{n-1,m-1}|\,f_{m}(t-|x-y|,y,p)\Bigg )= C\bigg(I_{1}+I_{2}+I_{3}\bigg).
\end{eqnarray*}
For $I_{3}$ we use that
\begin{eqnarray}
I_{3}&\leq& C\Delta\int\frac{dy}{|x-y|} \frac{|\delta F_{n-1,m-1}|(t-|x-y|,y)}{(1+|t-|x-y||+|y|)^{3}} \nonumber\\
&\leq& C\Delta\|\delta F_{n-1,m-1}\|_{3/4}\int
\frac{dy}{|x-y|} (1+|t-|x-y||+|y|)^{-19/4}\nonumber\\
&\leq& C\Delta \|\delta F_{n-1,m-1}\|_{3/4}(1+|t|+|x|)^{-1}(1+|t-|x||)^{-7/4}.\label{I3}
\end{eqnarray}
Here we used (\ref{trick}) with $\beta\equiv 2R$.
To estimate $I_{1}$ and $I_{2}$ in a proper way we need to carry out a factor $\|\delta F_{n-1,m-1}\|_{3/4}$.
To this purpose we notice that, by (\ref{vlasovn}),
\[
\partial_{t}\delta f_{n,m} +\widehat{p}\cdot\partial_{x}\delta f_{n,m}+K_{n-1}\cdot\partial_{p}\delta
f_{n,m}=-\delta K_{n-1,m-1}\cdot\partial_{p}f_{m}.
\]
Integrating along the characteristics of the Vlasov equation we get
\[
\delta f_{n,m}(t,x,p)=-\int_{0}^{t}(\delta E_{n-1,m-1}+\widehat{P}_{n}\wedge\delta B_{n-1,m-1})
\cdot\partial_{p}f_{m}(\tau,X_{n}(\tau),P_{n}(\tau))d\tau.
\]
>From the previous equation, inequality (\ref{estimdensn}) and the estimate $|X_n(\tau)|\leq R+a|\tau|$ we deduce 
\begin{equation}\label{estimdeltafn}
|\delta f_{n,m}(t,x,p)|\leq C\Delta\|\delta F_{n-1,m-1}\|_{3/4}(1+|t|)^{1/4}.
\end{equation}
Hence
\begin{eqnarray*}
I_{1}&\leq& C\Delta\|\delta F_{n-1,m-1}\|_{3/4}\int\frac{dy}{|x-y|^{2}}(1+|t-|x-y||+|y|)^{-11/4}\\
&\leq& C\Delta\|\delta F_{n-1,m-1}\|_{3/4}(1+|t|+|x|)^{-1}(1+|t-|x||)^{-3/4},
\end{eqnarray*}
\begin{eqnarray*}
I_{2}&\leq& C\Delta\|\delta F_{n-1,m-1}\|_{3/4}\int\frac{dy}{|x-y|}(1+|t-|x-y||+|y|)^{-19/4}\\ 
&\leq& C\Delta\|\delta F_{n-1,m-1}\|_{3/4}(1+|t|+|x|)^{-1}(1+|t-|x||)^{-7/4}.
\end{eqnarray*}
Adding the various estimates we get
\begin{displaymath}
(1+|t|+|x|)(1+|t-|x||)^{-3/4}|\delta E_{n,m}(t,x)|\leq C\Delta\|\delta F_{n-1,m-1}\|_{3/4}.
\end{displaymath}
An identical estimate holds for $\delta B_{n,m}$ and therefore we finally get
\begin{equation}
\|\delta F_{n,m}\|_{3/4}\leq C\Delta\|\delta F_{n-1,m-1}\|_{3/4}.
\end{equation}
If the initial data are small enough in order that $C\Delta<1$, then $F_{n}$ 
is a Cauchy sequence in the norm $\|\,\cdot\,\|_{3/4}$ and so it converges uniformly and the limit function $F=(E,B)$ 
satisfies 
\begin{equation}\label{quasiestimate1}
|F(t,x)|\leq C\Delta(1+|t|+|x|)^{-1}(1+|t-|x||)^{-3/4}.
\end{equation} \prfe

By (\ref{estimdeltafn}), the sequence $f_{n}(t,x,p)$ converges
uniformly with respect to $(t,x,p)\in [-T,T]\times\mathbb{R}^{3}_{x}\times\mathbb{R}^{3}_{p}$, for all $T>0$.
The limit function $(f,F)$ of the sequence $(f_{n},F_{n})$ is a continuous solution of RVM$_{\mathrm{ret}}$.  
Moreover, substituting (\ref{quasiestimate1}) into the second integral in the right hand side of (\ref{repfield0}), 
we find that $E(t,x)$ satisfies the estimate 
\[
|E(t,x)|\leq C\Delta (I_{2}^{3}+I_{1}^{19/4})\leq C\Delta(1+|t|+|x|)^{-1}(1+|t-|x||)^{-1}.
\]
The same is true for the magnetic field and so (\ref{estimate1}) is proved. 
\begin{Corollary}\label{estimdeltacharn}
The following inequalities hold for all $t\in \mathbb{R}$ and $(x,p)\in \Xi(t)$:
\begin{equation}\label{estimdeltafn2}
|\delta f_{n,m}(t,x,p)|\leq (1+|t|)^{1/4}q_{n,m},
\end{equation}
\begin{equation}\label{estdeltaXn}
|\delta X_{n,m}(0)|\leq(1+|t|)q_{n,m},
\end{equation}
\begin{equation}
|\delta P_{n,m}(0)|\leq q_{n,m},\label{estdeltaPn}
\end{equation}
where $q_{n,m}\to 0,$ as $n,m\to\infty$.
\end{Corollary}
\noindent\textit{Proof: }(\ref{estimdeltafn2}) follows by (\ref{estimdeltafn}). 
To prove (\ref{estdeltaXn}), (\ref{estdeltaPn}) we use that, by means of (\ref{charn}), say for $0\leq s\leq t$,
\[
|\delta X_{n,m}(s)|\leq\int_{s}^{t}d\tau\,|\delta\widehat{P}_{n,m}(\tau)|d\tau\leq C\int_{s}^{t}d\tau\,|\delta P_{n,m}(\tau)|.
\]
Moreover by the known $C^{1}$ bounds and the Cauchy property of $F_n$ in the norm $\|\,\cdot\,\|_{3/4}$, 
\begin{eqnarray*}
|\delta P_{n,m}(s)|&\leq&\int_{s}^{t}d\tau\,|K_{n-1}(\tau,X_{n}(\tau),P_{n}(\tau))-K_{m-1}(\tau,X_{m}(\tau),P_{m}(\tau))|\\
&\leq& \int_{s}^{t}d\tau\,|K_{n-1}(\tau,X_{n}(\tau),P_{n}(\tau))-K_{n-1}(\tau,X_{m}(\tau),P_{n}(\tau))|\\
&&+\int_{s}^{t}d\tau\,|K_{n-1}(\tau,X_{m}(\tau),P_{n}(\tau))-K_{n-1}(\tau,X_{m}(\tau),P_{m}(\tau))|\\
&&+\int_{s}^{t}d\tau\,|K_{n-1}(\tau,X_{m}(\tau),P_{m}(\tau))-K_{m-1}(\tau,X_{m}(\tau),P_{m}(\tau))|\\
&\leq& q_{n,m}+C\int_{s}^{t} d\tau\,(1+\tau)^{-11/4}|\delta X_{n,m}(\tau)|+C\int_{s}^{t}d\tau\,(1+\tau)^{-2}|\delta P_{n,m}(\tau)|.
\end{eqnarray*}
Combining the last two inequalities we get
\[
|\delta P_{n,m}(s)|\leq q_{n,m}+C\int_{s}^{t}d\tau[(\tau-s)(1+\tau)^{-11/4}+(1+\tau)^{-2}]|\delta P_{n,m}(\tau)|.
\]
Hence, by the Gronwall lemma:
\begin{equation}\label{estdeltacharn}
|\delta P_{n,m}(s)|\leq q_{n,m},\quad |\delta X_{n,m}(s)|\leq (1+t-s)q_{n,m},
\end{equation}
which concludes the proof. \prfe

\subsection*{Step 3: Convergence in the $C^{1}$ norm}
In this step we will prove the convergence of the sequence $DF_{n}$ with respect to the norm $\|\,\cdot\,\|_{1}$
(which again is not optimal but sufficient for our purpose).
\begin{Proposition}\label{convergence1}
For properly small initial data the sequence $DF_{n}$ converges in the norm $\|\,\cdot\,\|_1$.
\end{Proposition}
\noindent\textit{Proof:} Put $\delta Df_{n,m}=Df_{n}-Df_{m}$ and $\delta DE_{n,m}=DE_{n}-DE_{m}$. By (\ref{repderfield0}) we have 
\begin{eqnarray*}
\delta DE_{n,m}&=&\int dp \int dy\,\frac{\mathfrak{b}_{1}(\omega,p)}{|x-y|^{3}}\delta f_{n,m}(t-|x-y|,y,p)\\
&&+\int dp\int dy\,\frac{\mathfrak{b}_{2}(\omega,p)}{|x-y|^{2}}(f_{n}K_{n-1}-f_{m}K_{m-1})(t-|x-y|,y,p)\\
&&+\int dp \int dy\,\frac{\mathfrak{b}_{3}(\omega,p)}{|x-y|}[D(K_{n-1}f_{n})-D(K_{m-1}f_{m})](t-|x-y|,y,p)\\
&=&I_{1}+I_{2}+I_{3}.
\end{eqnarray*}
For $I_{2}$ we write 
\begin{eqnarray*}
|I_{2}(t,x)|&\leq& C\int dp\int \frac{dy}{|x-y|^{2}}|F_{n-1}|\,|\delta f_{n,m}|(t-|x-y|,y,p)\\
&&+C\int dp\int \frac{dy}{|x-y|^{2}}|\delta F_{n-1,m-1}|f_{m}(t-|x-y|,y,p)\\
&\leq& I_{2}^{19/4}q_{n,m}\leq q_{n,m}(1+|t|+|x|)^{-1}(1+|t-|x||)^{-11/4},
\end{eqnarray*}
where we used the estimate (\ref{estimdeltafn2}) and the Cauchy property of $F_{n}$ in the norm $\|\,\cdot\,\|_{3/4}$.
The integral $I_{1}$ is further split as follows:
\[
I_{1}=\int_{|x-y|\leq 1}dy\cdots + \int_{|x-y|>1} dy\cdots.
\]
For the second integral we have, by (\ref{estimdeltafn2}),
\[
\int_{|x-y|>1} dy\cdots\leq q_{n,m}\int_{|x-y|>1} \frac{dy}{|x-y|^{3}}(1+|t-|x-y||+|y|)^{-11/4}.
\]
The integral in the right hand side of the previous expression corresponds to the integral $II^{11/4}$ which has been estimated in 
the proof of lemma \ref{basiclemma2} (cf. (A.3) in appendix). The result is (see (A.4)) 
\begin{equation}\label{temp}
\int_{|x-y|>1} dy\cdots\leq q_{n,m}(1+|t|+|x|)^{-1}(1+|t-|x||)^{-3/2}.
\end{equation}
For the first part of the integral $I_{1}$, we have, by the same argument following eq. (A.2) in appendix,
\begin{equation}\label{temp2}
\int_{|x-y|\leq 1}dy\cdots\leq C\int_{t-1}^{t}\frac{\|\delta D f_{n,m}(\tau)\|_{\infty}}{(1+|\tau|+|x|)^{3}}.
\end{equation}
We will prove afterwards that
\begin{equation}\label{estdeltaDfn}
|\delta Df_{n,m}(t,x,p)|\leq (1+|t|)[q_{n,m}+C\Delta\|\delta DF_{n-1,m-1}\|_1].
\end{equation}
Hence substituting into (\ref{temp2}) and adding to (\ref{temp}) we get
\[
|I_{1}|\leq (1+|t|+|x|)^{-1} (1+|t-|x||)^{-1}(q_{n,m}+C\Delta\|\delta DF_{n-1,m-1}\|_1).
\]
For $I_{3}$ we expand the integrand function as
\begin{eqnarray*}
D(K_{n-1}f_{n})-D(K_{m-1}f_{m})&=&(DK_{n-1})\delta f_{n,m}+(Df_{m})\delta K_{n-1,m-1}\\
&&+f_{m}\delta DK_{n-1,m-1}+K_{n-1}\delta Df_{n,m}
\end{eqnarray*}
and therefore, after some straightforward estimates,
\[
|I_{3}|\leq (1+|t|+|x|)^{-1}(1+|t-|x||)^{-1}(q_{n,m}+C\Delta\|\delta DF_{n-1,m-1}\|_1).
\]
Summing up the various estimates we get
\[
(1+|t|+|x|)(1+|t-|x||)|\delta DE_{n,m}|\leq q_{n,m}+C\Delta\|\delta DF_{n-1,m-1}\|_1
\]
and so, by the analogous estimate for any other first derivative of $F_{n}$, we conclude
\[
\|\delta DF_{n,m}\|_1\leq q_{n,m}+C\Delta\|\delta DF_{n-1,m-1}\|_1.
\]
For properly small initial data, the previous inequality implies that $DF_{n}$ is a Cauchy sequence in the norm $\|\,\cdot\,\|_1$ 
and so that it converges uniformly. Therefore, $F$ is a $C^{1}$ function and satisfies:
\begin{equation}\label{quasiestimate2}
|DF(t,x)|\leq C\Delta(1+|t|+|x|)^{-1}(1+|t-|x||)^{-1}.
\end{equation}
Let us prove now the inequality (\ref{estdeltaDfn}), say for $t>0$. By (\ref{densn}) we have
\begin{eqnarray}
|\delta Df_{n,m}(t,x,p)|&\leq& |\partial_{x}\fzero(X_{n},P_{n})||\delta DX_{n,m}|+
|\partial_{p}\fzero(X_{n},P_{n})||\delta  DP_{n,m}|\nonumber\\
&&+|DX_{m}||\partial_{x}\fzero(X_{n},P_{n})-\partial_{x}\fzero(X_{m},P_{m})|\nonumber\\
&&+|DP_{m}||\partial_{p}\fzero(X_{n},P_{n})-\partial_{p}\fzero(X_{m},P_{m})|\nonumber\\
&\leq& C\Delta(|\delta DX_{n,m}|+|\delta DP_{n,m}|+|\delta X_{n,m}|+|\delta P_{n,m}|),\label{estdeltaDfn2}
\end{eqnarray}
evaluation of the characteristics at $s=0$ being understood.
By (\ref{charn}):
\begin{equation}\label{deltaDXn}
|\delta DX_{n,m}(s)|\leq \int_{s}^{t}d\tau\,\big(|\delta DP_{n,m}(\tau)|+|\delta P_{n,m}(\tau)|\big), 
\end{equation}
\begin{equation}\label{deltaDPn}
|\delta DP_{n,m}(s)|\leq \int_{s}^{t}d\tau\,|D[K_{n-1}(\tau,X_{n}(\tau),\widehat{P}_{n}(\tau))]-
D[K_{m-1}(\tau,X_{m}(\tau),\widehat{P}_{m}(\tau))]|.
\end{equation}
The integrand function in (\ref{deltaDPn}) is expanded as follows:
\begin{eqnarray*}
&&D[K_{n-1}(\tau,X_{n}(\tau),\widehat{P}_{n}(\tau))]-D[K_{m-1}(\tau,X_{m}(\tau),\widehat{P}_{m}(\tau))]\\
&&=\partial_{x}E_{n-1}(\tau,X_{n})\delta DX_{n,m}+DX_{m}[\partial_{x}E_{n-1}(\tau,X_{n})-\partial_{x}E_{m-1}(\tau,X_{m})]\\
&&\quad+B_{n-1}(\tau,X_{n})\wedge\delta D\widehat{P}_{n,m}+D\widehat{P}_{m}\wedge [B_{n-1}(\tau,X_{n})-B_{m-1}(\tau,X_{m})]\\
&&\quad+\widehat{P}_{n}\wedge\partial_{x}B_{n-1}(\tau,X_{n})\delta DX_{n,m}+\delta\widehat{P}_{x,m}\wedge\partial_{x}B_{n-1}(\tau,X_{n})DX_{m}\\
&&\quad+\widehat{P}_{m}\wedge DX_{m}[\partial_{x}B_{n-1}(\tau,X_{n})-\partial_{x}B_{m-1}(\tau,X_{m})].
\end{eqnarray*}
Using the known bounds on $F_{n}$ and $DF_{n}$ we get 
\begin{eqnarray}
|\delta DP_{n,m}(s)|&\leq& q_{n,m}+C\int_{s}^{t}d\tau\,(1+\tau)^{-11/4}|\delta DX_{n,m}(\tau)|
+C\int_{s}^{t}d\tau\,(1+\tau)^{-2}|\delta DP_{n,m}|\nonumber\\
&&+C\int_{t}^{s}d\tau\,|B_{n-1}(\tau,X_{n}(\tau))-B_{m-1}(\tau,X_{m}(\tau))|\nonumber\\
&&+C\int_{t}^{s}d\tau\,|DF_{n-1}(\tau,X_{n}(\tau))-DF_{m-1}(\tau,X_{m}(\tau))|.\label{estdeltaDPn}
\end{eqnarray}
Now we substitute (\ref{deltaDXn}) into (\ref{estdeltaDPn}) and use
\begin{eqnarray*}
&&|B_{n-1}(\tau,X_{n}(\tau))-B_{m-1}(\tau,X_{m}(\tau))|\\
&&\leq  |B_{n-1}(\tau,X_{n}(\tau))-B_{m-1}(\tau,X_{n}(\tau))|
+|B_{m-1}(\tau,X_{n}(\tau))-B_{m-1}(\tau,X_{m}(\tau))|\\
&&\leq [(1+\tau)^{-7/4}+(1+t-\tau)(1+\tau)^{-11/4}]\,q_{n,m},
\end{eqnarray*}
\begin{eqnarray*}
&&|DF_{n-1}(\tau,X_{n}(\tau))-DF_{m-1}(\tau,X_{m}(\tau))|\\
&&\leq |DF_{n-1}(\tau,X_{n}(\tau))-DF_{m-1}(\tau,X_{n}(\tau))|
+|DF_{m-1}(\tau,X_{n}(\tau))-DF_{m-1}(\tau,X_{m}(\tau))|\\
&&\leq C(1+\tau)^{-2}\|\delta F_{n-1,m-1}\|_1
+[(1+t-\tau)(1+\tau)^{-11/4}]q_{n,m},
\end{eqnarray*}
which follow by the known bounds on the first and second order derivatives, the second of (\ref{estdeltacharn}) and the Cauchy property 
of $F_{n}$ in the norm $\|\,\cdot\,\|_{3/4}$. In this way we get
\begin{eqnarray*}
|\delta DP_{n,m}(s)|&\leq& q_{n,m}+C\|\delta DF_{n-1,m-1}\|_1\\
&&+\int_{s}^{t}d\tau\,[(1+\tau)^{-2}+(1+t-\tau)(1+\tau)^{-11/4}]|\delta DP_{n,m}(\tau)|
\end{eqnarray*}
and so, by the Gronwall lemma,
\[
|\delta DP_{n,m}|\leq (q_{n,m}+C\|\delta DF_{n-1,m-1}\|_1).
\]
Thus by (\ref{deltaDXn}),
\begin{equation}
|\delta DX_{n,m}|\leq (q_{n,m}+C\|\delta DF_{n-1,m-1}\|_1)(1+t-s).
\end{equation}
Taking $s=0$ and substituting into (\ref{estdeltaDfn2}), the estimate (\ref{estdeltaDfn}) follows after using (\ref{estdeltaXn}) and 
(\ref{estdeltaPn}). \prfe

By means of (\ref{estdeltaDfn}), $Df_{n}$ converges uniformly in $x,p$ and pointwise in $t$. The same argument permits to prove that even 
the $p$-derivatives of $f_{n}$ satisfy this property and therefore the limit function $(f,F)$ is $C^{1}$.
Substituting (\ref{quasiestimate2}) into the last integral of (\ref{repderfield0}), the estimate (\ref{estimate2}) is proved by using again 
lemmas \ref{basiclemma1} and \ref{basiclemma2}.
This concludes the proof of theorem \ref{main}.

\bigskip
\noindent
{\bf Acknowledgments:}
The author acknowledges many useful discussions with his PhD advisor, Dr. Alan D. Rendall. Support by the European HYKE network 
(contract HPRN-CT-2002-00282) is also acknowledged.

\section*{Appendix}
\renewcommand{\theequation}{A.\arabic{equation}}
\setcounter{equation}{0}
\subsection*{Proof of lemma 4}
We will use repeatedly lemma 7 of \cite{GS4}, which we rewrite below in a form more suitable to our case.

\noindent\textbf{Lemma A} \textit{For any function $g\in C^{0}(\mathbb{R}^{2})$, $a>0$, $b\in (a,+\infty]$ and $n\in\mathbb{N}$:}
\[
\int_{a\leq |x-y|\leq b}\frac{dy}{|x-y|^{n}}\,g(t-|x-y|,|y|)=\frac{2\pi}{|x|}\int_{t-b}^{t-a}\frac{d\tau}{(t-\tau)^{n-1}}
\int_{||x|-t+\tau|}^{|x|+t-\tau}
d\lambda\,g(\tau,\lambda)\lambda.
\]

\subsubsection*{Estimate on $I_{1}^{q}$ and $I_{2}^{q}$ for $t\leq 0$}
For $t\leq 0$ we have $|t-|x-y||=-t+|x-y|$ and by lemma A we have: 
\begin{eqnarray*}
I_{1}^{q}(t,x)&=&
\int\frac{dy}{|x-y|}(1-t+|x-y|+|y|)^{-q}\\
&=&\frac{2\pi}{|x|}\int_{-\infty}^{t}d\tau\int_{||x|-t+\tau|}^{|x|+t-\tau}\frac{\lambda}
{(1+\lambda-\tau)^{q}}d\lambda\\
&=&\frac{2\pi}{|x|}\int_{-\infty}^{t-|x|}\cdots+\frac{2\pi}{|x|}\int_{t-|x|}^{t}\cdots=A+B.
\end{eqnarray*}
For A we use
\begin{eqnarray*}
A&=&\frac{2\pi}{|x|}\int_{-\infty}^{t-|x|}d\tau\int_{t-|x|-\tau}^{t+|x|-\tau}\frac{\lambda}{(1+\lambda-\tau)^{q}}
d\lambda\\
&\leq&\frac{C}{|x|}\int_{-\infty}^{t-|x|}d\tau\int_{t-|x|-\tau}^{t+|x|-\tau}\frac{d\lambda}{(1+\lambda-\tau)^{q-1}}\\
&\leq& C\int_{-\infty}^{t-|x|}\frac{d\tau}{(1-\tau)^{q-1}}\leq\frac{C}{(1-t+|x|)^{q-2}}.
\end{eqnarray*}
For B we use
\begin{eqnarray*}
B&=&\frac{2\pi}{|x|}\int_{t-|x|}^{t}d\tau\int_{|x|-t+\tau}^{|x|+t-\tau}\frac{\lambda }{(1+\lambda-\tau)^{q}}d\lambda\\
&\leq&\frac{C}{|x|(1-t+|x|)^{q-1}}\int_{t-|x|}^{t} (t-\tau)d\tau\\
&\leq&\frac{C}{(1-t+|x|)^{q-2}}.
\end{eqnarray*}
For $n=2$, $t\leq 0$, we write, again using lemma A,
\begin{eqnarray*}
I_{2}^{q}(t,x)&=&\int\frac{dy}{|x-y|^{2}}(1-t+|x-y|+|y|)^{-q}\\
&=&\frac{2\pi}{|x|}\int_{-\infty}^{t}\frac{d\tau}{t-\tau}\int_{||x|-t+\tau|}^{|x|+t-\tau}\frac{\lambda}
{(1+\lambda-\tau)^{q}}d\lambda\\
&=&\frac{2\pi}{|x|}\int_{-\infty}^{t-|x|}\cdots+\frac{2\pi}{|x|}\int_{t-|x|}^{t}\cdots=A+B.
\end{eqnarray*}
For A we use
\begin{eqnarray*}
A&=&
\frac{2\pi}{|x|}\int_{-\infty}^{t-|x|}\frac{d\tau}{t-\tau}\int_{t-|x|-\tau}^{t+|x|-\tau}
\frac{\lambda}{(1+\lambda-\tau)^{q}}d\lambda\\
&\leq&\frac{C}{|x|}\int_{-\infty}^{t-|x|}d\tau\Bigg (\frac{t+|x|-\tau}{t-\tau}\Bigg
)\int_{t-|x|-\tau}^{t+|x|-\tau}\frac{d\lambda}{(1+\lambda-\tau)^{q}}\\
&\leq&\frac{C}{|x|}\int_{-\infty}^{t-|x|}d\tau\frac{|x|}{(1-\tau)^{q}}=\frac{C}{(1-t+|x|)^{q-1}}.
\end{eqnarray*}
For B we use 
\begin{eqnarray*}
B&=&\frac{2\pi}{|x|}\int_{t-|x|}^{t}\frac{d\tau}{t-\tau}\int_{|x|-t+\tau}^{|x|+t-\tau}\frac{\lambda}
{(1+\lambda-\tau)^{q}}d\lambda\\
&\leq&\frac{C}{|x|}\int_{t-|x|}^{t}\frac{d\tau}{t-\tau}
\int_{|x|-t+\tau}^{|x|+t-\tau}\frac{1}{(1+\lambda-\tau)^{q-1}}d\lambda\\
&\leq&\frac{C}{|x|(1-t+|x|)^{q-1}}\int_{t-|x|}^{t}\frac{d\tau}{t-\tau}\int_{|x|-t+\tau}^{|x|+t-\tau}d\lambda\\
&\leq&\frac{C}{(1-t+|x|)^{q-1}}.
\end{eqnarray*}

\subsubsection*{Estimate on $I_{1}^{q}$ for $t>0$}
We split $I_{1}^{q}$ as follows:
\begin{eqnarray*}
I_{1}^{q}(t,x)&=&\int_{|x-y|\leq t}\frac{dy}{|x-y|}(1+t-|x-y|+|y|)^{-q}\\
&&+\int_{|x-y|\geq t}\frac{dy}{|x-y|}(1-t+|x-y|+|y|)^{-q}=I_{1A}^{q}+I_{1B}^{q}.
\end{eqnarray*}
By using lemma A we have
\begin{eqnarray*}
&&I_{1A}^{q}=\frac{2\pi}{|x|}\int_{0}^{t}d\tau\int_{||x|-t+\tau|}^{|x|+t-\tau}d\lambda\frac{\lambda}{(1+\tau+\lambda)^{q}},\\
&&I_{1B}^{q}=\frac{2\pi}{|x|}\int_{-\infty}^{0}d\tau\int_{||x|-t+\tau|}^{|x|+t-\tau}d\lambda\frac{\lambda}{(1-\tau+\lambda)^{q}}.
\end{eqnarray*}
Now define
\[
(t-|x|)_{+}=\left\{\begin{array}{ll}
t-|x| & \textrm{if } t-|x|>0\\
0 & \textrm{if } t-|x|\leq 0
\end{array}\right.
\]
\[
(t-|x|)_{-}=\left\{\begin{array}{ll}
t-|x| & \textrm{if } t-|x|<0\\
0 & \textrm{if } t-|x|\geq 0
\end{array}\right.
\]
and split the preceding integrals as follows:
\begin{eqnarray*}
I_{1A}^{q}&=&\frac{2\pi}{|x|}\int_{0}^{(t-|x|)_{+}}d\tau\int_{t-|x|-\tau}^{|x|+t-\tau}d\lambda\frac{\lambda}{(1+\tau+\lambda)^{q}}\\
&&+\frac{2\pi}{|x|}\int_{(t-|x|)_{+}}^{t}d\tau\int_{|x|-t+\tau}^{|x|+t-\tau}d\lambda\frac{\lambda}{(1+\tau+\lambda)^{q}}=
I_{1A\alpha}^{q}+I_{1A\beta}^{q},
\end{eqnarray*}
\begin{eqnarray*}
I_{1B}^{q}&=&\frac{2\pi}{|x|}\int_{-\infty}^{(t-|x|)_{-}}d\tau\int_{t-|x|-\tau}^{|x|+t-\tau}d\lambda\frac{\lambda}{(1-\tau+\lambda)^{q}}\\
&&+\frac{2\pi}{|x|}\int_{(t-|x|)_{-}}^{0}d\tau\int_{|x|-t+\tau}^{|x|+t-\tau}d\lambda\frac{\lambda}{(1-\tau+\lambda)^{q}}=
I_{1B\alpha}^{q}+I_{1B\beta}^{q}.
\end{eqnarray*}
Thus, finally
\begin{equation}
I_{1}^{q}(t,x)=I_{1A\alpha}^{q}+I_{1A\beta}^{q}+I_{1B\alpha}^{q}+I_{1B\beta}^{q}.
\end{equation}

\noindent \textit{Estimate for $I_{1A\alpha}^{q}$}
\begin{eqnarray*}
I_{1A\alpha}^{q}&\leq&
\frac{C}{|x|}\int_{0}^{(t-|x|)_{+}}d\tau\int_{t-|x|-\tau}^{|x|+t-\tau}\frac{d\lambda}{(1+\tau+\lambda)^{q-1}}\\
&\leq&
\frac{C}{|x|}\int_{0}^{(t-|x|)_{+}}\frac{d\tau}{(1+t-|x|)^{q-3}}\int_{t-|x|-\tau}^{|x|+t-\tau}\frac{d\lambda}{(1+\tau+\lambda)^{2}}\\
&\leq& \frac{C(t-|x|)_{+}}{(1+t-|x|)^{q-2}(1+t+|x|)}\leq C(1+t+|x|)^{-1}(1+|t-|x||)^{-q+3}.
\end{eqnarray*}

\noindent \textit{Estimate for $I_{1A\beta}^{q}$}
\begin{eqnarray*}
I_{1A\beta}^{q}&\leq&\frac{C}{|x|}\int_{(t-|x|)_{+}}^{t}d\tau\int_{|x|-t+\tau}^{|x|+t-\tau}\frac{d\lambda}{(1+\tau+\lambda)^{q-1}}\\
&\leq&\frac{C}{|x|}\int_{(t-|x|)_{+}}^{t}d\tau\frac{t-\tau}{(1-t+|x|+2\tau)^{q-2}(1+t+|x|)}\\
&\leq& \frac{C(t-(t-|x|)_+)}{|x|(1+|t|+|x|)}\int_{(t-|x|)_+}^td\tau\,(1-t+|x|+2\tau)^{2-q}\\
&\leq& \frac{C}{1+|t|+|x|}\Bigg (\frac{(1+|x|+(t-|x|)_+)^{q-3}}{(1+t+|x|)^{q-3}(1-t+|x|+2(t-|x|)_+)^{q-3}}\Bigg )\\
&\leq& C(1+t+|x|)^{-1}(1+|t-|x||)^{-q+3}.
\end{eqnarray*}

\noindent \textit{Estimate for $I_{1B\alpha}^{q}$}
\begin{eqnarray*}
I_{1B\alpha}^{q}&\leq&
\frac{C}{|x|}\int_{-\infty}^{(t-|x|)_{-}}d\tau\int_{t-|x|-\tau}^{|x|+t-\tau}\frac{d\lambda}{(1-\tau+\lambda)^{q-1}}\\
&\leq& C\int_{-\infty}^{(t-|x|)_{-}}\frac{d\tau}{(1+t-|x|-2\tau)^{q-2}(1+t+|x|-2\tau)}\\
&\leq& C (1+t+|x|)^{-1}(1+t-|x|-2(t-|x|)_{-})^{-q+3}\\
&\leq& C(1+t+|x|)^{-1}(1+|t-|x||)^{-q+3}.
\end{eqnarray*}

\noindent \textit{Estimate for $I_{1B\beta}^{q}$}
\begin{eqnarray*}
I_{1B\beta}^{q}&\leq& \frac{C}{|x|}\int_{(t-|x|)_{-}}^{0}\frac{(t-\tau) d\tau}{(1-t+|x|)^{q-2}(1+t+|x|-2\tau)}\\
&\leq&\frac{C}{|x|}\frac{(t-(t-|x|)_{-})|(t-|x|)_{-}|}{(1-t+|x|)^{q-2}(1+t+|x|)}\\
&\leq& C(1+t+|x|)^{-1}(1+|t-|x||)^{-q+3}.
\end{eqnarray*}

\subsubsection*{Estimate on $I_{2}^{q}$ for $t>0$}
The integral $I_{2}^{q}$ is split as $I_{1}^{q}$ in (A.1), namely
\[
I_{2}^{q}(t,x)=I_{2A\alpha}^{q}+I_{2A\beta}^{q}+I_{2B\alpha}^{q}+I_{2B\beta}^{q},
\]
where, using lemma A,
\begin{eqnarray*}
&&I_{2A\alpha}^{q}=\frac{2\pi}{|x|}\int_{0}^{(t-|x|)_{+}}\frac{d\tau}{t-\tau}\int_{t-|x|-\tau}^{|x|+t-\tau}d\lambda\frac{\lambda}
{(1+\tau+\lambda)^{q}},\\
&&I_{2A\alpha}^{q}=\frac{2\pi}{|x|}\int_{(t-|x|)_{+}}^{t}\frac{d\tau}{t-\tau}\int_{|x|-t+\tau}^{|x|+t-\tau}d\lambda\frac{\lambda}
{(1+\tau+\lambda)^{q}},\\
&&I_{2B\alpha}^{q}=\frac{2\pi}{|x|}\int_{-\infty}^{(t-|x|)_{-}}\frac{d\tau}{t-\tau}\int_{t-|x|-\tau}^{|x|+t-\tau}d\lambda\frac{\lambda}
{(1-\tau+\lambda)^{q}},\\
&&I_{2B\beta}^{q}=\frac{2\pi}{|x|}\int_{(t-|x|)_{-}}^{0}\frac{d\tau}{t-\tau}\int_{|x|-t+\tau}^{|x|+t-\tau}d\lambda\frac{\lambda}
{(1-\tau+\lambda)^{q}}.
\end{eqnarray*}
Since the argument to estimate the preceding integrals is very similar to the one used for $I_{1}^{q}$, we just show how to
estimate $I_{2A\alpha}^{q}$. 
\begin{eqnarray*}
I_{2A\alpha}^{q}&\leq&\frac{C}{|x|}\int_{0}^{(t-|x|)_{+}}d\tau\frac{|x|+t-\tau}{t-\tau}\int_{t-|x|-\tau}^{|x|+t-\tau}\frac{d\lambda}
{(1+\lambda+\tau)^{q}}\\
&\leq& \frac{C}{|x|}\Bigg ( 1+\frac{|x|}{t-(t-|x|)_{+}}\Bigg )\int_{0}^{(t-|x|)_{+}}d\tau\frac{|x|}{(1+t-|x|)^{q-1}(1+t+|x|)}\\
&\leq& \frac{C(t-|x|)_{+}}{(1+t+|x|)(1+t-|x|)^{q-1}}\leq C(1+t+|x|)^{-1}(1+|t-|x||)^{-q+2}.
\end{eqnarray*}

\subsection*{Proof of lemma 5}
We divide the integral $I(t,x)$ in two parts as follows:
\begin{equation}
I=\int_{|x-y|\leq 1}dy\cdots + \int_{|x-y|>1} dy\cdots.
\end{equation}
Following \cite{GS2}, we rewrite the first integral as
\begin{eqnarray*}
\int_{|x-y|\leq 1}dy\cdots &=&\int_{t-1}^{t}d\tau\int dp\int_{|\omega|=1} \mathfrak{b}_{1}(\omega,p)\frac{g(\tau,x+(t-\tau)\omega,p)}{t-\tau}d\omega\\
&=&\int_{t-1}^{t}d\tau\int dp\int_{|\omega|=1} \mathfrak{b}_{1}(\omega,p)\frac{g(\tau,x+(t-\tau)\omega,p)-g(\tau,x,p)}{t-\tau}d\omega,
\end{eqnarray*}
where the property (\ref{zeroaverage}) has been used. Then
\begin{eqnarray*}
\int_{|x-y|\leq 1}dy\cdots&\leq& C_{*}\sup_{t-1\leq\tau\leq t}\|Dg(\tau)\|_{\infty}\int_{t-1}^{t}\frac{d\tau}{(1+|\tau|+|x|)^{3}}\\
&\leq& C_{*}\|Dg\|_{\infty}(1+|t|+|x|)^{-3}\\
&\leq& C_{*}\|Dg\|_{\infty}(1+|t|+|x|)^{-1}(1+|t-|x||)^{-2}.
\end{eqnarray*}
For the second part of $I$ we write
\[
\int_{|x-y|>1} dy\cdots\leq C_{*}\|g\|_{\infty}\int_{|x-y|>1}\frac{dy}{|x-y|^{3}}(1+|t-|x-y||+|y|)^{-3}=C_{*}\|g\|_{\infty}II(t,x).
\]
Since an integral similar to $II(t,x)$ needs to be estimated to prove proposition \ref{convergence1}, we will treat the more general case
\begin{equation}
II^{q}(t,x)=\int_{|x-y|>1}\frac{dy}{|x-y|^{3}}(1+|t-|x-y||+|y|)^{-q},\quad q>2.
\end{equation}
We will prove that
\begin{equation}
II^{q}(t,x)\leq C (1+|t|+|x|)^{-1} (1+||t-|x||)^{-q+5/4}.
\end{equation}
We start by splitting $II^{q}(t,x)$ as follows:
\[
II^{q}(t,x)=\int_{1<|x-y|\leq 1+|t-|x||}\cdots + \int_{|x-y|> 1+|t-|x||}\cdots=II^{q}_{A}+II^{q}_{B}.
\]
For $II_{B}^{q}$ we use
\[
II_{B}^{q}\leq \frac{I_{2}^{q}(t,x)}{(1+|t-|x||)}\leq C(1+|t|+|x|)^{-1}(1+|t-|x||)^{-q+1}.
\]
The estimate on $II_{A}^{q}$ for $t\leq 0$ is 
\begin{eqnarray*}
II_{A}^{q}&\leq& \int_{1<|x-y|\leq 1-t+|x|}\frac{dy}{|x-y|^{3}}(1-t+|x-y|+|y|)^{-q}\\
&\leq&\frac{C}{(1-t+|x|)^{q}}\int_{1\leq |x-y|\leq 1-t+|x|}dy |x-y|^{-3}\\
&\leq&\frac{C\log(1-t+|x|)}{(1-t+|x|)^{q}}\leq C(1+|t|+|x|)^{-1}(1+|t-|x||)^{-q+\frac{5}{4}}.
\end{eqnarray*}
The estimate on $II_{A}^{q}$ for $t>0$ requires a more careful analysis.

\subsubsection*{Estimate on $II_{A}^{q}$ for $t>0, |x|\leq 1$}
For $t\leq 1$, $II_A^q$ is dominated by the same integral extended over $\{1\leq |x-y|\leq 3\}$ and so the estimate is straightforward. 
For $t\geq 1$ we have 
$t-|x|\geq 0$ and so we may split $II_{A}^{q}$ as follows:
\begin{eqnarray*}
II_{A}^{q}&=&\int_{1\leq |x-y|\leq t}\frac{dy}{|x-y|^{3}}(1+t-|x-y|+|y|)^{-q}\\
&&+\int_{t\leq |x-y|\leq 1+t-|x|}\frac{dy}{|x-y|^{3}}(1-t+|x-y|+|y|)^{-q}\\
&=&II^{q}_{A1}+II_{A2}^{q}.
\end{eqnarray*}
Using lemma A we have
\[
I_{A1}^{q}=
\frac{2\pi}{|x|}\int_{0}^{t-1}\frac{d\tau}{(t-\tau)^{2}}\int_{||x|+\tau-t|}^{|x|+t-\tau}d\lambda\frac{\lambda}{(1+\tau+\lambda)^{q}}.
\]
Since $t-1\leq t-|x|$, then $|x|+\tau-t\leq 0$ and we have
\begin{eqnarray*}
II^{q}_{A1}&\leq&\frac{C}{|x|}\int_{0}^{t-1}\frac{d\tau}{t-\tau}\frac{|x|+t-\tau}{t-\tau}\frac{1}{(1+t-|x|)^{q-2}}
\int_{t-|x|-\tau}^{|x|+t-\tau}\frac{d\lambda}{(1+\tau+\lambda)^{2}}\\
&\leq& C(1+t-|x|)^{-q+1}(1+t+|x|)^{-1}\int_{0}^{t-1}d\tau (t-\tau)^{-1} \\
&\leq& C\frac{\log t}{(1+t-|x|)^{q-1}}(1+t+|x|)^{-1} \\
&\leq& C(1+t+|x|)^{-1}(1+|t-|x||)^{-q+\frac{5}{4}},
\end{eqnarray*}
since $t\geq 1$ and $t-|x|\geq 0$. 

For $II_{A2}^{q}$ we write
\begin{eqnarray*}
II^q_{A2}&=&\frac{2\pi}{|x|}\int_{|x|-1}^{0}\frac{d\tau}{(t-\tau)^{2}}\int_{t-|x|-\tau}^{|x|+t-\tau}d\lambda\frac{\lambda}
{(1-\tau+\lambda)^{q}}\\
&\leq&\frac{C}{|x|}\int_{|x|-1}^{0}\frac{d\tau}{t-\tau}\frac{|x|}{(1+t-|x|-2\tau)^{q-1}(1+t+|x|-2\tau)}\\
&\leq& C\frac{\log t+\log (1+t-|x|)}{(1+t+|x|)(1+t-|x|)^{q-1}}\\
&\leq& C(1+t+|x|)^{-1}(1+|t-|x||)^{-q+\frac{5}{4}}.
\end{eqnarray*}
Since in the following the details are very similar, they will be omitted.

\subsubsection*{Estimate on $II_{A}^{q}$ for $|x|> 1, 0<t\leq 1$}
In this case we have $t-|x|\leq 0$ and $|x-y|> 1\geq t$
and so
\begin{eqnarray*}
II_{A}^{q}&=&\int_{1< |x-y|\leq 1-t+|x|}\frac{dy}{|x-y|^{3}}(1-t+|x-y|+|y|)^{-q}\\
&=&\frac{2\pi}{|x|}\int_{2t-1-|x|}^{t-1}\frac{d\tau}{(t-\tau)^{2}}\int_{||x|+\tau-t|}^{|x|+t-\tau}d\lambda\frac{\lambda}
{(1-\tau+\lambda)^{q}}.
\end{eqnarray*}
Since $2t-1-|x|\leq t-|x|\leq t-1$, we split the last integral as follows:
\begin{eqnarray*}
II_{A}^{q}&=&\frac{2\pi}{|x|}\int_{2t-1-|x|}^{t-|x|}\frac{d\tau}{(t-\tau)^{2}}\int_{t-|x|-\tau}^{|x|+t-\tau}
d\lambda\frac{\lambda}{(1-\tau+\lambda)^{q}}\\
&&+\frac{2\pi}{|x|}\int_{t-|x|}^{t-1}\frac{d\tau}{(t-\tau)^{2}}\int_{|x|+\tau-t}^{|x|+t-\tau}d\lambda\frac{\lambda}
{(1-\tau+\lambda)^{q}}=II^{q}_{A1}+II_{A2}^{q},
\end{eqnarray*}
and each component is estimated as before.

\subsubsection*{Estimate on $II_{A}^{q}$ for $|x|> 1, t> 1$}
\textit{Case $t-|x|\geq 0$}

\noindent Since $1+t-|x|<t$, we have
\begin{eqnarray*}
II_{A}^{q}&=&\int_{1< |x-y|\leq 1+t-|x|}\frac{dy}{|x-y|^{3}}(1+t-|x-y|+|y|)^{-q}\\
&=&\frac{2\pi}{|x|}\int_{|x|-1}^{t-1}\frac{d\tau}{(t-\tau)^{2}}\int_{||x|+\tau-t|}^{|x|+t-\tau}d\lambda\frac{\lambda}
{(1+\tau+\lambda)^{q}}.
\end{eqnarray*}
We further consider separately the regions $\frac{1}{2}(t+1)<|x|\leq t$ and $1<|x|\leq \frac{1}{2}(t+1)$. In the first case one has
$|x|-1>t-|x|$ and therefore $II_{A}^{q}$ reduces to
\[
II_{A}^{q}=\frac{2\pi}{|x|}\int_{|x|-1}^{t-1}\frac{d\tau}{(t-\tau)^{2}}\int_{|x|+\tau-t}^{|x|+t-\tau}d\lambda\frac{\lambda}
{(1+\tau+\lambda)^{q}},
\] 
which is estimated as before. For $1<|x|\leq \frac{1}{2}(t+1)$ we write
\begin{eqnarray*}
II_{A}^{q}&=&\frac{2\pi}{|x|}\int_{|x|-1}^{t-|x|}\frac{d\tau}{(t-\tau)^{2}}\int_{t-|x|-\tau}^{|x|+t-\tau}
d\lambda\frac{\lambda}{(1+\tau+\lambda)^{q}}\\
&&+\frac{2\pi}{|x|}\int_{t-|x|}^{t-1}\frac{d\tau}{(t-\tau)^{2}}\int_{|x|+\tau-t}^{|x|+t-\tau}d\lambda\frac{\lambda}
{(1+\tau+\lambda)^{q}}=II^{q}_{A1}+II_{A2}^{q}
\end{eqnarray*}
and each component is estimated as before.

\vspace{0.3 cm}
\noindent\textit{Case $t-|x|<0$}

\noindent For $|x|\leq 2t-1$ we write
\begin{eqnarray*}
II_{A}^{q}&=&\int_{1\leq |x-y|\leq 1-t+|x|}\frac{dy}{|x-y|^{3}}(1+t-|x-y|+|y|)^{-q}\\
&=&\frac{2\pi}{|x|}\int_{2t-|x|-1}^{t-1}\frac{d\tau}{(t-\tau)^{2}}\int_{|x|+\tau-t}^{|x|+t-\tau}d\lambda\frac{\lambda}
{(1+\tau+\lambda)^{q}},
\end{eqnarray*}
where we used that $t-|x|< 2t-|x|-1$. For $|x|\geq 2t-1$ we write
\begin{eqnarray*}
II_{A}^{q}&=&\int_{1\leq |x-y|\leq t}\frac{dy}{|x-y|^{3}}(1+t-|x-y|+|y|)^{-q}\\
&&+\int_{t\leq |x-y|\leq 1-t+|x|}\frac{dy}{|x-y|^{3}}(1-t+|x-y|+|y|)^{-q}\\
&=&\frac{2\pi}{|x|}\int_{0}^{t-1}\frac{d\tau}{(t-\tau)^{2}}\int_{|x|+\tau-t}^{|x|+t-\tau}d\lambda\frac{\lambda}
{(1+\tau+\lambda)^{q}}\\
&&+\frac{2\pi}{|x|}\int_{2t-|x|-1}^{0}\frac{d\tau}{(t-\tau)^{2}}\int_{|x|+\tau-t}^{|x|+t-\tau}d\lambda\frac{\lambda}
{(1-\tau+\lambda)^{q}}.
\end{eqnarray*}
The usual argument applies to estimate all the above integrals and concludes the proof of lemma 5.


\end{document}